\documentclass{jfm}
\usepackage{graphicx}
\usepackage{amsmath}
\usepackage{mathtools}
\usepackage{epstopdf, epsfig}
\usepackage{dcolumn}%
\usepackage{bm}%
\usepackage{multirow}
\usepackage{caption}
\usepackage{subcaption}
\usepackage{float}
\usepackage{placeins}
\usepackage[dvipsnames]{xcolor}
\definecolor{magenta}{RGB}{255,0,255}
\definecolor{green}{RGB}{0,153,76.5}
\usepackage{tikz}
\usetikzlibrary{shapes}
\usetikzlibrary{plotmarks}
\usepackage{pgfplots}
\usepackage{comment}
\usepackage{hyperref}
\usepackage[symbol]{footmisc}

\usepackage{threeparttable}

\newcommand{\blackline}{\raisebox{2pt}{\tikz{\draw[-,black,solid,line width = 0.75pt](0.,0mm) -- (5mm,0mm)}}}

\newcommand{\Grayline}{\raisebox{2pt}{\tikz{\draw[-,Gray,solid,line width = 0.75pt](0.,0mm) -- (5mm,0mm)}}}
\newcommand{\redline}{\raisebox{2pt}{\tikz{\draw[-,red,solid,line width = 0.75pt](0.,0mm) -- (5mm,0mm)}}}

\newcommand{\Limegreenline}{\raisebox{2pt}{\tikz{\draw[-,ForestGreen,solid,line width = 0.75pt](0.,0mm) -- (5mm,0mm)}}}

\newcommand{\blueline}{\raisebox{2pt}{\tikz{\draw[-,blue,solid,line width = 0.75pt](0.,0mm) -- (5mm,0mm)}}}
\newcommand{\greenline}{\raisebox{2pt}{\tikz{\draw[-,green,solid,line width = 0.75pt](0.,0mm) -- (5mm,0mm)}}}
\newcommand{\magentaline}{\raisebox{2pt}{\tikz{\draw[-,magenta,solid,line width = 0.75pt](0.,0mm) -- (5mm,0mm)}}}

\newcommand{\hollowcircleline}{\raisebox{0pt}{\tikz{\draw[black,solid,line width = 0.5pt](2.5mm,0) circle [radius=0.75mm];\draw[-,black,dotted,line width = 0.5pt](0.,0mm) -- (5mm,0mm)}}}

\newcommand{\hollowcirclesolidline}{\raisebox{0pt}{\tikz{\draw[black,solid,line width = 0.5pt](2.5mm,0) circle [radius=0.75mm];\draw[-,black,solid,line width = 0.5pt](0.,0mm) -- (5mm,0mm)}}}

\newcommand{\hollowsquaresolidline}{\raisebox{0pt}{\tikz{\draw[black,solid,line width = 0.5pt](2.mm,0) rectangle (3.5mm,1.5mm);\draw[-,black,solid,line width = 0.5pt](0.,0.8mm) -- (5.5mm,0.8mm)}}}
\newcommand{\hollowtrainglesolidline}{\raisebox{0pt}{\tikz{\draw[black,solid,line width = 0.5pt](2.mm,0) -- (3.5mm,0) -- (2.75mm,1.5mm) -- (2.mm,0);\draw[-,black,solid,line width = 0.5pt](0.,0.8mm) -- (5.5mm,0.8mm)}}}
\newcommand{\hollowstarsolidline}{\raisebox{0pt}{\tikz{\draw[black,solid,line width = 0.5pt](2.mm,0.8mm) -- (3.5mm,0.8mm); \draw[black,solid,line width = 0.5pt](2.75mm,1.5mm) -- (2.75mm,0);\draw[black,solid,line width = 0.5pt](2.325mm,1.25mm) -- (3.25mm,0.25mm);\draw[black,solid,line width = 0.5pt](2.325mm,0.25mm) -- (3.25mm,1.25mm);\draw[-,black,solid,line width = 0.5pt](0.,0.8mm) -- (5.5mm,0.8mm)}}}
\newcommand{\solidcircleline}{\raisebox{0pt}{\tikz{\draw[fill=black,solid,line width = 0.5pt](2.5mm,0) circle [radius=0.75mm];\draw[-,black,dotted,line width = 0.5pt](0.,0mm) -- (5mm,0mm)}}}

\shorttitle{Lift and drag near a wall}
\shortauthor{N. I. K. Ekanayake \textit{et} al.}

\title{Lift and drag forces acting on a particle moving with zero slip velocity near a wall}

\author{Nilanka. I. K. Ekanayake\aff{1},
  Joseph D. Berry\aff{1}, Anthony D. Stickland\aff{1}, David E. Dunstan \aff{1}, Ineke L. Muir\aff{2}, Steven K. Dower\aff{2}
 \and Dalton J. E. Harvie\aff{1}\corresp{\email{daltonh@unimelb.edu.au}}}

\affiliation{\aff{1}Department of Chemical Engineering, The University of Melbourne, Victoria 3010 Australia
\aff{2}
CSL, Bio21 Molecular Science and Biotechnology Institute, Victoria 3052, Australia}

\begin{document}

\maketitle

\begin{abstract}
The lift and drag forces acting on a small, neutrally-buoyant spherical particle in a single-wall-bounded linear shear flow are examined via numerical computation. The effects of shear rate are isolated from those of slip by setting the particle velocity equal to the local fluid velocity (zero slip), and examining the resulting hydrodynamic forces as a function of separation distance. In contrast to much of the previous numerical literature, low shear Reynolds numbers are considered ($10^{-3} \lesssim Re_{\gamma} \lesssim 10^{-1}$). This shear rate range is relevant when dealing with particulate flows within small channels, for example particle migration in microfluidic devices being used or developed for the biotech industry.  We demonstrate a strong dependence of both the lift and drag forces on shear rate.  Building on previous theoretical $Re_{\gamma} \ll 1$ studies, a wall-shear based lift correlation is proposed that is applicable when the wall lies both within the inner and outer regions of the disturbed flow. Similarly, we validate an improved drag correlation that includes higher order terms in wall separation distance that more accurately captures the drag force when the particle is close to, but not touching, the wall. Application of the new correlations shows that the examined shear based lift force is as important as the previously examined slip based lift force, highlighting the need to account for shear when predicting the near-wall movement of neutrally-buoyant particles.  %
\end{abstract}


\section{Introduction}
Particles moving in a sheared flow exhibit cross-stream migration and cluster at different equilibrium distances across channels \citep{Segre62}. In the absence of inter-particle collisions (ie, a dilute suspension), this clustering is primarily a result of hydrodynamic lift forces that cause particles to migrate in a direction normal to the flow. Lift forces can be important in (for example) cancer-detecting and cell sorting applications used in novel microfluidic devices and for flow cytometry \citep{DiCarlo09}. In the context of blood flow, lift forces may aid in the separation of platelets from red blood cells, causing the formation of a cell free layer (CFL) adjacent to blood vessel walls. CFL development is crucial for blood clot formation, as the platelet concentration is increased in the CFL, enhancing the hemostatic coagulation mechanisms required to repair damaged vessel walls \citep{Leiderman}.

Hydrodynamic lift forces are inertial in origin, and hence reduce to zero for rigid particles in the Stokes (low Reynolds number) limit \citep{Bretherton}. At finite Reynolds numbers and in unbounded linear flows, rigid particles experience lift when moving relative to the undisturbed fluid velocity (ie, at a finite slip velocity) either when there is finite shear within the fluid, or particle rotation relative to the fluid \citep{Saffman,Rubinow}. In bounded flows (ie, near a wall), particles moving at finite Reynolds numbers experience an additional lift force even in the absence of slip.  Additionally, the drag force, which is often much higher in magnitude than the lift force, is also affected by the presence of a wall, causing neutrally-buoyant particles moving at small but finite Reynolds numbers tangential to a wall to lag the fluid \citep{Goldman2}.  Few studies have analysed the lift and drag forces acting on a particle due to shear and/or rotation near a wall when there is no particle slip, however knowledge of these forces is required to model the cross-stream migration of neutrally-buoyant particles, relevant to applications including those described above. 
Hence the motivation for this study.

The lift force acting on a rigid sphere was first examined by \citet{Saffman}, who considered a buoyant particle moving at a finite slip velocity in an unbounded linear shear flow. The model considered the inertial effects of the far field disturbed flow on the lift force, at low slip and shear Reynolds numbers ($Re_{\text{slip}} \ll 1$, $Re_{\gamma} \ll 1$), where
\begin{equation}
    Re_{\text{slip}} = \frac{u_{\text{slip}}a}{\nu},~~~~~~
    Re_{\gamma} = \frac{\gamma a^2}{\nu},
    \label{Reynoldsnumers}
\end{equation}
and $u_{\text{slip}}$, $a$, $\nu$ and $\gamma$ are the particle slip velocity, particle radius, fluid kinematic viscosity and fluid shear rate, respectively. Unlike shear free flows, in which the Stokes length scale ($L_{\text{S}} ={\nu}/{u_\text{slip}}$) is used to separate the inner and outer regions, an additional length scale known as the Saffman length scale ($L_{\text{G}} = \sqrt{{\nu}/{\gamma}}$) is considered for linear shear flows.  Generally the boundary of the inner and outer region in a linear shear flow field is located at min$(L_{\text{G}},L_{\text{S}})$. Saffman's lift is an ``outer region" model that solves the velocity field in the region far away from the particle where both viscous and inertial effects are significant. Specifically, an Oseen-type equation is solved via a matched asymptotic expansion method and the flow disturbance due to the particle is determined by treating the particle as a point force. This model neglects the significance of viscous effects in the ``inner region" closer to the particle. In addition to the low Reynolds number condition, Saffman's lift model is valid only when the inertial effects due to shear rate are much higher than the inertial effects generated by slip velocity ($\epsilon = \sqrt{Re_{\gamma}}/Re_{\text{slip}}\gg 1$ or equivalently ${L_\text{G}} \ll {L_\text{S}}$). \citet{Asmolov90} and \citet{McLaughlin1} independently relaxed this constraint on $\epsilon$ presenting an unbounded lift model for comparable inertial effects ($\sqrt{Re_{\gamma}} \sim Re_{\text{slip}}$) that reduces to Saffman's result when $\epsilon \rightarrow \infty$. The above unbounded lift models reduce to zero if either $Re_{\text{slip}}$ or $Re_{\gamma}$ is zero.

Particle rotation also affects the forces experienced by a particle.  A freely translating particle in an unbounded shear flow rotates at a rate corresponding to a zero torque condition and hence, experiences a rotational lift force \citep{Rubinow}. However, in a linear shear flow, \citet{Saffman} illustrated that the lift force due to rotation is less by an order of magnitude than that due to the slip-shear of a freely rotating sphere translating with a small slip. Thus, most unbounded lift models only consider a non-rotating particle \citep{McLaughlin1,Asmolov90}.

\begin{figure}
    \centerline{\includegraphics[width=0.95\linewidth]{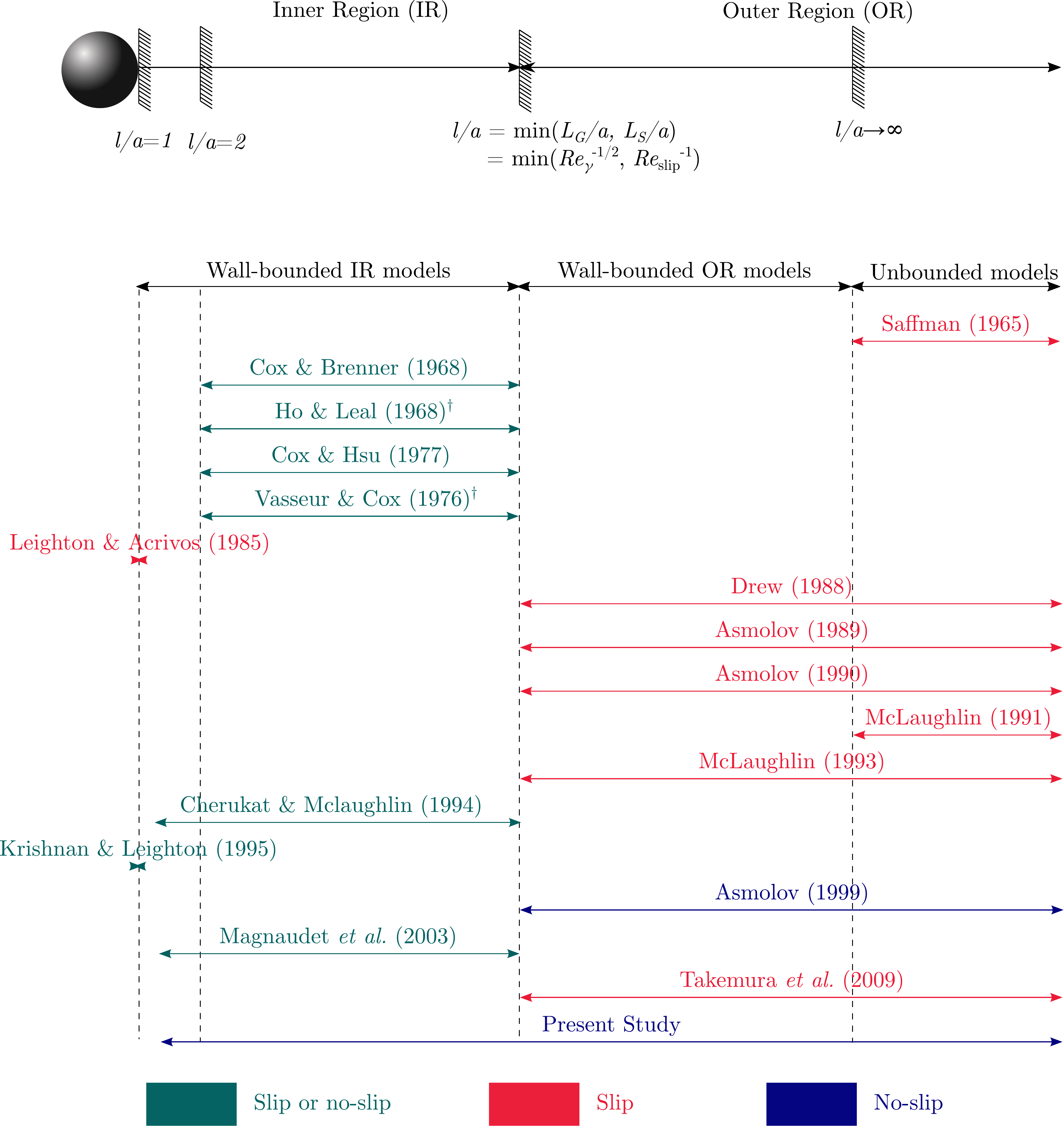}}
    \caption{Analytical lift force models available for linear shear flows illustrating the range of applicability based on dimensionless wall distance ($l/a$) and on ``slip" and ``no-slip" conditions. The lift coefficient of ``slip" models scales with $Re_\text{slip}$ and $Re_{\gamma}$ whereas the lift coefficient of ``no-slip" models scales only with $Re_\text{slip}$. $^\text{\textdagger}$Models bounded by two walls.}
    \label{LiftModels}
\end{figure}

In addition to the slip-shear and rotation effects, the presence of a wall near a particle also affects the hydrodynamic forces experienced by that particle.  These wall-induced inertial forces have been discussed in a number of `bounded flow' studies. Figure \ref{LiftModels} summarises the available analytical lift models for both these bounded and unbounded flows. As illustrated, in bounded flow models the walls can reside in either the inner or outer regions of the disturbed flow surrounding the particle.

Most theoretical wall-bounded lift models require the wall to lie within the inner region, $l \ll$ min$(L_{\text{G}},L_{\text{S}})$, where $l$ is the distance between the particle centre and the wall. Additionally, these models require $Re_{\text{slip}}, Re_{\gamma} \ll 1$. Inner region models are applicable for particles having both zero and finite slip velocities, with studies generally using the term `neutrally-buoyant particles' to refer to zero slip results, and `buoyant particles' to refer to finite slip velocity results. \citet{CoxBrenner} were the first to obtain an implicit expression for the forces induced on arbitrary shaped particles by a wall lying in the inner region. Simplifying this model, \citet{CoxHsu} derived a closure expression for the migration velocity ($u_\text{m}$) of a rigid spherical particle freely translating in single wall-bounded flow. The equivalent lift force was also calculated using Stokes law. Predictions of $u_\text{m}$ were obtained for both zero slip (`neutrally-buoyant') and slip (`buoyant') particle velocities. The model is valid only when the separation distance is large compared to the sphere radius ($l/a \gg 1$), with the equation for the force having a leading order term proportional to ${l/a}$. Using the method of reflections, \citet{HoLeal} explicitly calculated the lift force acting on a freely rotating, neutrally-buoyant particle (zero slip) bounded by two flat walls for both Couette and Poiseuille flows at low Reynolds numbers. This inner region study accounted for the effects of both walls and the flow curvature present in Poiseuille flow. For Couette flow the model predicted that particles migrate towards the channel centre due to a wall-shear lift force that is greatest near the walls but zero at the channel centre. The predicted particle migration in Couette flow agrees substantially with the experimental observations of \citet{Halow}. Later, \citet{VasseurCox} extended the theoretical work of \citet{CoxHsu} to analyse the migration velocity of spherical particles in linear flows bounded by two flat plates, with both plates residing in the inner region of the particle. The predicted migration velocities of \citet{VasseurCox} for neutrally-buoyant particles agreed well with the results of \citet{HoLeal} in the channel centre, but deviated significantly in the vicinity of the wall.  As explained in \citet{VasseurCox}, this deviation is due to poor convergence of the numerical computation in the \citet{HoLeal} solution, particularly when the sphere is close to a wall.  The same model predictions of \citet{VasseurCox} agreed well with the asymptotic behaviour suggested by the single wall \citet{CoxHsu} model when the particle is near one of the walls.

Several other inner region studies \citep{Leighton,Krishnan,CherukatMcLaughlin} have performed lift force analyses when the particle is almost in contact with the wall ($l/a \gtrsim 1$). In contrast to other inner region studies where the particle is usually treated as a point-force, these studies consider higher-order contributions to the flow disturbances induced by the particle, accounting for the finite size of the particle. The limiting case of a stationary particle touching the wall \citep{Leighton} and a freely translating and rotating particle almost in contact with a wall \citep{Krishnan} in a linear shear flow were studied using lubrication analysis to predict asymptotic lift coefficient values. Accounting for the finite size of the particle, the lift force variation with wall distance was analysed by \citet{CherukatMcLaughlin}, down to a minimum separation distance of ${l/a} = 1.1$. The extrapolated results for a particle touching the wall agreed well with latter studies \citep{Krishnan,Leighton} for both fixed and freely translating cases. More recently, \citet{Magnaudet1} derived inner region solutions for a spherical bubble and a rigid particle moving freely near a wall and proposed lift correlations valid for $l/a \gtrsim 1$. 

In all the inner region studies, the lift force on both rotating and non-rotating particles is obtained by coupling the two flow disturbances that originate from particle slip and fluid shear, the stokeslet and stresslet, respectively, in a non-linear manner. These inner region lift models present the force as
\begin{equation}
    {F_{L}^{\ast}} = \frac{F_{\text{L}} \rho}{\mu^2} = C_\text{L,1}Re_{\gamma}^2+C_\text{L,2}Re_{\gamma}Re_{\text{slip}}+C_\text{L,3}Re_{\text{slip}}^2
    \label{eq:lift1}
\end{equation}
where the three lift coefficients ($C_\text{L1},C_\text{L2},C_\text{L3}$) are functions of $a/l$. The first term on the right hand side of Eq. (\ref{eq:lift1}) originates from the disturbance induced by the presence of the wall in a shear flow field, and is independent of slip. The second term depends on both the slip velocity and shear rate. The last term, independent of the shear rate, is due to the stokeslet generated from slip velocities in a quiescent fluid next to a wall.

 As the inner region models require a particle to be close to a wall, they cannot be used to predict unbounded results as ${l/a}$ becomes large.  Instead, a few authors have investigated the effect of walls lying in outer region of the disturbed flow, that is, under conditions where $Re_{\gamma}, Re_\text{slip} \ll 1$. These outer region wall-bounded models use the method of matched asymptotic expansions to solve the singular perturbation problem by treating the particle as a point force. Unlike inner region models, outer region models correctly predict the unbounded results as ${l/L_\text{G}} \rightarrow \infty$. Outer region studies analysing a particle sedimenting in a stagnant fluid ($Re_\gamma = 0$) have presented lift correlations as functions of $l/L_\text{S}$ \citep{VasseurCox2, Takemura2003, Takemura, Shi2}. Similarly, linear shear flows bounded by a single wall were examined by \citet{Drew88} and \citet{Asmolov89} for the condition $\epsilon \gg 1$. Later, \citet{Asmolov90} and \citet{McLaughlin2} extended their previous analysis to single wall-bounded linear flows valid for all $\epsilon$. The latter two studies considered a non-rotating, buoyant spherical particle with a finite slip velocity and the calculated lift coefficients were tabulated/plotted as a function of $l/L_\text{G}$ and $\epsilon$. Using a similar approach, \citet{Asmolov99} evaluated the lift force coefficient for a freely rotating neutrally-buoyant particle in a linear shear flow bounded by a single wall located in the outer region. Recently, \citet{TakemuraMagnaudet1} obtained a semi-empirical lift coefficient model based on the McLaughlin results which recovers the correct asymptotic behaviours when the particle is located both near and far from the wall, for $\epsilon \gg 1$ and $\epsilon \ll 1$ conditions. However, as this correlation only captures slip-shear based lift, it predicts zero lift for a zero slip (`neutrally-buoyant') particle.

In this study we are particularly interested in the lift force on a slip free, or neutrally-buoyant, particle. When the wall is located close to the particle (in the inner region of the disturbed flow field), the lift force can be evaluated from Eq. (\ref{eq:lift1}) by using $Re_{\text{slip}}=0$. Noting that $F_\text{L}$ is non-dimensionalised by $\rho a^4 \gamma^2 (={Re_{\gamma}}^2\mu^2/\rho)$, we see that the net lift force is determined by the first lift coefficient ($C_\text{L,1}$). The available formulations for $C_\text{L,1}$ are summarised in table \ref{InnerLiftSummary}. The listed inner region models are valid only for low shear and slip Reynolds numbers ($Re_{\gamma}, Re_\text{slip} \ll 1$), with the lift coefficients ($C_\text{L,1}$) asymptoting to a finite value of $\sim~1.9$ for a non-rotating sphere and ${\sim~1.8}$ for a freely rotating sphere as the particle approaches the wall \citep{CoxHsu,Krishnan,CherukatMcLaughlin}. Conversely, when ${l/a}$ becomes large, although the inner region models predict a finite lift coefficient ($C_\text{L,1}$) with increasing separation distance (for finite but low $Re_{\gamma}$), as the wall moves from the inner to outer region the results become invalid.  Conversely, most models that are valid when the wall resides in the outer region \citep{McLaughlin2,Asmolov90,Asmolov89, TakemuraMagnaudet1} predict a zero lift force for a no-slip particle as the force scales with $Re_\text{slip}$ in addition to $Re_{\gamma}$. As an exception, \citet{Asmolov99} derived an outer region wall-bounded lift model for neutrally-buoyant (no-slip) particles that gives a lift force that scales with $Re_{\gamma}$ and varies as a function of $l/L_\text{G}$. Nevertheless, no analytical model is able to predict the lift force on a zero slip (neutrally-buoyant) particle across both the inner and outer regions of the flow, as a function of the shear rate within the fluid.

\newcommand{\tagarray}{%
\mbox{}\refstepcounter{equation}%
$(\theequation)$%
}
\begin{table}
\begin{center}
  \resizebox{\textwidth}{!}{%
  \begin{tabular}{lccc}
    Study & $C_\text{L,1}$ & Comments & Eq. \\
    \hline
    \\
    \multirow{4}{*}{\citet{CoxHsu}} & $\dfrac{366\pi}{576}$  & Non-rotating, ${l/a} \gg 1$ & \multirow{2}{*}\quad\tagarray\label{Cox1}\\
    \\
    & $\dfrac{330\pi}{576}$ & Freely-rotating, ${l/a} \gg 1$ & \multirow{2}{*}\quad\tagarray\label{Cox2}\\
    \\
    \hline
    \\
    {\citet{CherukatMcLaughlin}} & $2.0069+1.0575\bigg(\dfrac{a}{l}\bigg)-2.4007\bigg(\dfrac{a}{l}\bigg)^{2}+1.3174\bigg(\dfrac{a}{l}\bigg)^{3}$ & Non-rotating, ${l/a} \gtrsim 1$ & \quad\tagarray\label{CM1}\\
    \\
    {\citet{CherukatMcLaughlinCorrection}} & $1.8065+0.89934\bigg(\dfrac{a}{l}\bigg)-1.961\bigg(\dfrac{a}{l}\bigg)^{2}+1.02161\bigg(\dfrac{a}{l}\bigg)^{3}$ & Freely rotating, ${l/a} \gtrsim 1$ & \quad\tagarray\label{CM2}\\
    \\
    \hline
    \\
    \multirow{2}{*}{\citet{Krishnan}} & $1.9680$ & Non-rotating, ${l/a}=1$ & \quad\tagarray\label{KL1}\\
    \\
    & $1.6988$ & Freely rotating, ${l/a}=1$&\quad\tagarray\label{KL2}\\
    \\
    \hline
    \\
    {\citet{Magnaudet1}} & $\dfrac{55\pi}{96}\bigg[1+\dfrac{9}{16}\bigg(\dfrac{a}{l}\bigg)\bigg]$ & Freely rotating, ${l/a} \gtrsim 1$&\quad\tagarray\label{MD1}\\
    \\
  \end{tabular}}
\end{center}
\caption{Wall-shear lift coefficients ($C_\text{L,1}$) of inner region studies}
 \label{InnerLiftSummary}
\end{table}

Drag, a force present in both Stokes and inertial flows, is equally as important as the lift force in predicting the migration of particles, as drag can induce slip, and in the presence of wall slip alone (or slip plus shear) can induce a lift force. In unbounded fluid flow, drag is only a function of the slip velocity of the particle, however the presence of a wall increases this slip-induced drag such that it is highest when the particle is in contact with the wall, and decays rapidly to the unbounded value as the separation distance increases. Analyses of wall-bounded slip-induced drag have been conducted for both Stokes \citep{Happel,Goldman1,Goldman2,Magnaudet1} and inertial flows \citep{VasseurCox2}. For Stokes flows, Faxen \citet{Happel} deduced a wall correction for the slip based drag resulting in higher order terms (in separation distance, $\mathcal{O}((a/l)^5)$) being included in the force expansion. Using a method of matched asymptotic expansion, \citet{VasseurCox2} suggested a slip based inertial correction term for Faxen's inner region drag model, as well as an outer region slip based drag model that gives the force as a function of $l/L_\text{G}$ for zero shear flows.

For linear shear flows, \citet{Magnaudet1} presented an additional contribution to the drag force due to wall-shear (independent of slip) giving the net drag force on a translating spherical particle with a wall in the inner region as
\begin{equation}
    {F_{D}^{\ast}} = \frac{F_{\text{D}} \rho}{\mu^2} = C_\text{D,1}Re_{\gamma}+C_\text{D,2}Re_{\text{slip}}
    \label{eq:drag1}
\end{equation}
Here $C_\text{D,1},$ is a function of only $(a/l)$ and $C_\text{D,2}$ is a function of both $(a/l)$ and $Re_{\text{slip}}$.  The first term in Eq. (\ref{eq:drag1}) represents the drag force due to shear (independent of slip), and is relevant for neutrally-buoyant particles in linear shear flows. For low shear rates \citet{Magnaudet1} gives $C_\text{D,1}$ as
\begin{equation}
C_\text{D,1}=\frac{15}{8}\pi\bigg(\dfrac{a}{l}\bigg)^{2}\bigg[1+\frac{9}{16}\bigg(\dfrac{a}{l}\bigg)\bigg]
\label{eq:dragCD1}
\end{equation}
This $C_\text{D,1}$ reduces rapidly to zero away from the wall \citep{Magnaudet1}. The second term $C_\text{D,2}$ in Eq. (\ref{eq:drag1}) accounts for the drag acting on the particle due to slip and asymptotes to the low Reynolds number Stokes drag value far from the wall. In comparison to the slip based drag coefficient $C_\text{D,2}$, only a few studies have examined the behaviour of the inertial shear based coefficient $C_\text{D,1}$. Note that a freely translating ($F_\text{D}^{\ast}=0$) neutrally-buoyant particle in a linear shear flow lags the fluid flow near the wall due to the negative slip generated by the wall shear drag, as explained by \citet{Goldman2} under Stokes flow conditions and by \citet{Magnaudet1} for finite inertial flow conditions. 

In contrast to the above theoretical works, most numerical studies that calculate the forces acting on a wall-bounded particle are for intermediate Reynolds numbers $Re_{\gamma}, Re_{\text{slip}} \sim 0.5 - 3\times10^2$. Among these, simulations on translating particles performed in quiescent flows are used to evaluate the effect of separation distance on both slip based lift and drag correlations \citep{Zeng2005}. Several other direct numerical studies examining hydrodynamic forces acting on a particle in a linear shear flow and with finite slip closer to the wall propose inertial corrections that are functions of both shear rate and slip velocity \citep{Zeng2009, Lee}. To perform inner region simulations, satisfying $Re_{\gamma}, Re_{\text{slip}} < 1$ conditions, one must employ large computational domains with high mesh refinement near the particle surface in order to capture the small inertial forces. These simulations require high computational power. To our knowledge no numerical results have been presented for small but finite inertial conditions $Re_\gamma$ or $Re_\text{slip} \leq \mathcal{O}(1)$, possibly due to these computational limitations, particularly for wall-bounded flows under zero slip conditions.

Here we consider a rigid spherical particle moving at the same velocity as the fluid (zero slip) in a linear shear flow tangential to a flat wall. The objective is to extend the $Re_\gamma \ll 1 $ results to low but finite Reynolds numbers, in the absence of slip, using results from a large number of well resolved numerical simulations. We use non-rotating as well as freely rotating spheres, for shear Reynolds numbers in the range of $10^{-3}$ to $10^{-1}$.  We express our results as new slip-free lift and drag coefficient correlations, and then use these results to examine how shear induced particle slip affects particle lift forces and migration velocities of neutrally-buoyant particles. Throughout this study both freely rotating and non-rotating particles are considered.

\section{Problem Specification}\label{sec:Problem Specification}

\begin{figure}
    \begin{subfigure}{0.495\linewidth}
    \centerline{\includegraphics[width=0.75\linewidth]{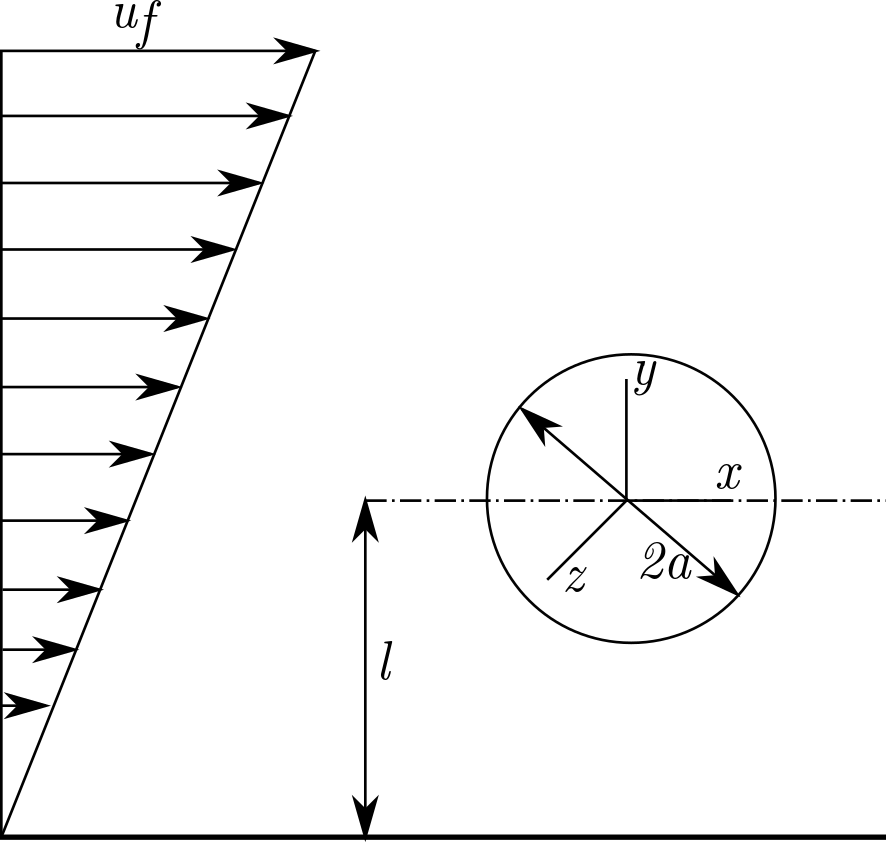}}
    \caption{}
    \label{Diagram1}
    \end{subfigure}
    \begin{subfigure}{0.495\linewidth}
    \centerline{\includegraphics[width=0.95\linewidth]{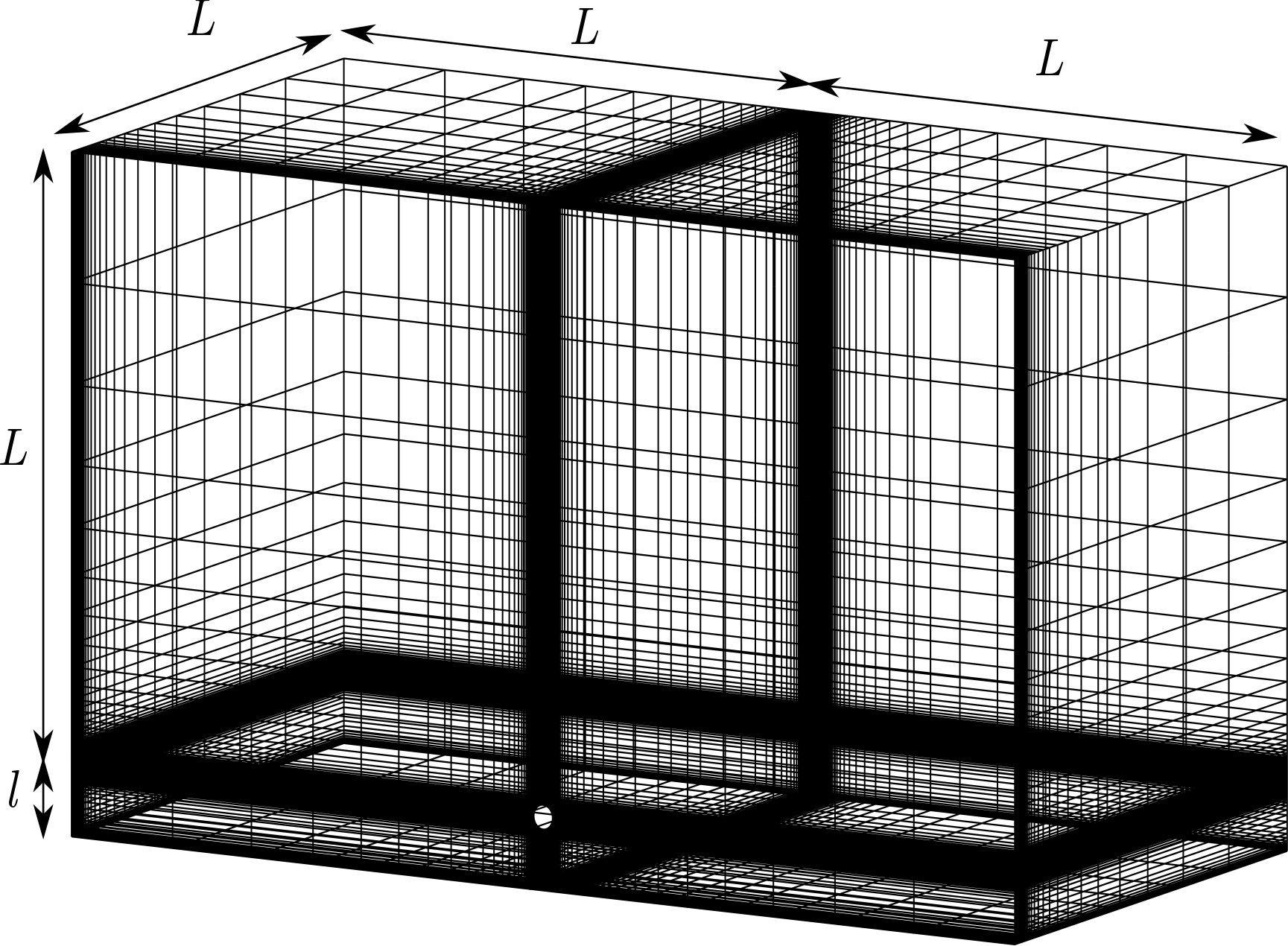}}
    \caption{}
    \label{Mesh}
    \end{subfigure}
\caption{(a) Schematic of a translating sphere of radius $a$ moving at velocity $u_\text{p}$ in a wall-bounded linear shear flow (b) The domain mesh for a particle located at $l/a=4$. }
\end{figure}

A neutrally-buoyant rigid sphere of radius $a$ is suspended in a linear shear flow with the origin of the Cartesian coordinate system located at the centre of the sphere (Figure \ref{Diagram1}). The coordinate unit vectors are ${\boldsymbol{e}_\text{x}}$, ${\boldsymbol{e}_{\text{y}}}$ and ${\boldsymbol{e}_\text{z}}$. A no-slip wall is placed at distance $(0, -l, 0)$ away from sphere centre. Outer boundaries are located at large distances $L(\gg l)$ away from the sphere centre to minimise any secondary boundary effects. For this study ${l/a}$ is varied from 1.2 - 9.5 to obtain the lift and drag force variation as a function of particle distance from the wall.

To obtain the forces generated due to wall and shear effects in the absence of relative motion between the particle and the fluid, the particle slip velocity ${\boldsymbol{u}_{\text{slip}}}$ is explicitly set to zero: That is,
\begin{align*}
  \boldsymbol{u_{\text{slip}}} &=\boldsymbol{u_{\text{p}}}-\boldsymbol{u_{\text{f}}}(y=0) = \boldsymbol{0}
\end{align*}
where ${\boldsymbol{u}_{\text{p}}}=u_\text{p}\boldsymbol{e_{\text{x}}}$ is the particle velocity and ${\boldsymbol{u}_{\text{f}}}$ is the undisturbed fluid velocity defined as
\begin{align*}
  \boldsymbol{u_{\text{f}}} &=\dot{\gamma}(y+l) \boldsymbol{e_{\text{x}}}
\end{align*}
Note that under this formulation the particle is constrained to translate only in the $x$ direction with particle velocity $u_\text{p}=\dot{\gamma}l$.

To determine the forces acting on the particle, we solve the steady-state Navier Stokes (N-S) equations in a frame of reference that moves with the particle \citep{Batchelor67}.  Specifically, we solve
\begin{gather}
\displaystyle{\boldsymbol{\nabla} \cdot \rho\boldsymbol{u'}} = 0\\
\displaystyle  {\boldsymbol{\nabla} \cdot (\rho\boldsymbol{u'}\boldsymbol{u'}+\boldsymbol{\sigma})} = 0
\label{NS}
\end{gather}
where $\boldsymbol{u'} = \boldsymbol{u} - \boldsymbol{u_{\text{p}}}$ and $\boldsymbol{u}$ is the local fluid velocity.  The boundary conditions used in the moving frame of reference are:
\begin{equation}
\boldsymbol{u'}=\left \{ \begin{array}{ll}  
\displaystyle \dot{\gamma}(y+l)\boldsymbol{e_{\text{x}}}-\boldsymbol{u_\text{p}} \ \ \ \  y=+\infty; y=-l; x,z=\pm\infty \\
\displaystyle \boldsymbol{\omega}\times\boldsymbol{r} \qquad\qquad\ \ \ \  {|\boldsymbol{ r}|}=a
\end{array}\right.
\label{BC2}
\end{equation}
where $\boldsymbol{r}$ is a radial displacement vector pointing from the sphere centre to the particle surface and $\boldsymbol{\omega}$ is the angular rotation of the particle.

The fluid is assumed to be Newtonian with a dynamic viscosity $\mu$ and density $\rho$. The total stress tensor ($\boldsymbol{\sigma}= p\boldsymbol{I}+\boldsymbol{\tau}$) is computed using the fluid pressure, $p$ and viscous stress tensor, $\boldsymbol{\tau} = -\mu(\boldsymbol{\nabla u'} + \boldsymbol{\nabla u'}^{\text{T}})$. The forces acting on the particle are evaluated by integrating the total stress contributions around the particle surface $S$:
\begin{equation}
    {\boldsymbol{F}_{\text{p}}} = -\int_S \boldsymbol{n}\cdot{\boldsymbol{\sigma}} dA
    \label{Force}
\end{equation}
Here $\boldsymbol{n} (=\boldsymbol{\hat{r}})$ is a unit normal vector directed out of the particle. The drag (${F_\text{D}} = {\boldsymbol{F}_{\text{p}}}\cdot\boldsymbol{e}_{\text{x}}$) and lift (${F_\text{L}}={\boldsymbol{F}_{\text{p}}}\cdot\boldsymbol{e}_{\text{y}} $) are defined as the fluid forces acting on the sphere in $+x$ and $+y$ directions, respectively.  Similarly the net torque $\boldsymbol{T}_{\text{p}}$ acting on the particle is evaluated using:
\begin{equation}
    {\boldsymbol{T}_{\text{p}}} = -\int_S \boldsymbol{r} \times {\boldsymbol{\sigma}}\cdot\boldsymbol{n} dA 
    \label{Torque}
\end{equation}
In this study, two cases are considered: in the first, the particle is constrained from rotating, whereas, in the second, the particle is allowed to freely rotate about the $\text{z}$ axis, at an angular velocity $\omega_\text{p}$. For the non-rotating (first) case all components of the rotation $\boldsymbol{\omega}$ are explicitly set to zero, whereas for the (second) freely rotating case the $z$ component of the net torque $\boldsymbol{T}_{\text{p}}$ is explicitly set to zero and the $z$ component of $\boldsymbol{\omega}$ ($\boldsymbol{\omega} \cdot \boldsymbol{e_{\text{z}}} = \omega_\text{p}$) is solved for as an unknown (with other components of $\boldsymbol{\omega}$ set to zero).  

Unless stated otherwise, the results in the remainder of this study are presented in non-dimensional form (indicated by an asterisk) using a length scale $a$, velocity scale $\gamma a$, and force scale $\mu^2/\rho$.

\section{Numerical approach}\label{sec:Numerical approach}

The system of equations is solved using the finite volume package \textit{arb} \citep{Harvie} over a non-uniform body-fitted structured mesh that is generated with \textit{gmsh} \citep{gmsh} (Figure \ref{Mesh}). The sphere surface is resolved with $N_{\text{p}}$ mesh points on each curved side length of a cubed-sphere. This results in $6(N_{\text{p}} - 1)^2$ cells on the sphere surface. $(N_{\text{p}}-1)$ number of inflation layers with an $\alpha$ geometric progression ratio around the sphere are used to capture gradients in velocity occurring near the particle wall. Outer boundaries, except the bottom wall located at $y=-l^{\ast}$, are placed at a distance $L^{\ast}$ from the sphere centre in all directions. To resolve the far field of the domain (excluding the inflation layers), $N_{\text{d}}$ points are used with a geometric progression $\beta$ expanding towards the domain boundaries. For larger separation distances, ${l^{\ast}} > 1.2$, ($N_{\text{w}}-1$) layers are introduced in the gap between the bounding box and the bottom wall, with a non-uniform progression of $\beta$ producing more refinement near the sphere and the bottom wall. The total cell count in the mesh is $N_{\text{t}}$.

The non-dimensional lift force magnitudes measured in this study are significantly low, $|F_{L}^{\ast}|~\sim~10^{-6}~- 10^{-1}$, and hence require a high resolution mesh around the sphere to accurately determine the lift coefficient. Further, the lift force is highly sensitive to the cell distribution around the particle, requiring perfect symmetry of the mesh in the $y$ direction to correctly ensure that the lift reduces to zero as inertial effects are reduced. For this reason the symmetry of the sphere surface mesh and its surrounding cells in the bounding box are enforced by employing a ``lego" type construction (Figure \ref{BlockMesh}) whereby groups of mesh `blocks' are copied, translated, rotated and joined to construct the full computational domain. For example, a single block mesh ($B1$) is copied and then rotated about the $x=0$ and $y=0$ axes to create identically discretised blocks that surround the sphere ($B1_\text{x},B1_\text{y}$ and $B1_\text{xy}$). Similarly, the upper and lower mesh domain blocks are built using two separate block mesh entities, $B2$ and $B3$ as illustrated in figure \ref{BlockMesh}, which are later rotated about $x=0$ axis. This mesh configuration ensures that we compute the zero lift force on a slip free particle under Stokes conditions down to $F_\text{L}^{\ast} \sim 10^{-15}$. The shear and slip Reynolds numbers considered in this study are sufficiently small to assume flow symmetry about the $z = 0$ plane. Hence a final domain size of [$-L, L$], [$L, -l$] and [$L, 0$] is used for simulations.

\subsection{Domain Size Dependency}\label{sec:Domain Size Dependency}
\begin{table}
\centering
    \begin{tabular}{ccccllllllll}
    \multirow[t]{3}{*}{${l^{\ast}}$} & \multirow[t]{3}{*}{$Re_{\gamma}$} & \multicolumn{2}{c}{Domain} & \multicolumn{4}{c}{Non-rotating} & \multicolumn{4}{c}{Freely-rotating} \\\\
    & & & & 
      \multicolumn{2}{c}{Lift} &
      \multicolumn{2}{c}{Drag} &
      \multicolumn{2}{c}{Lift} &
      \multicolumn{2}{c}{Drag}\\\\
    & & ${L^{\ast}}$  & $N_\text{t}$ & $C_{\text{L,1}}$ & 
    $\delta$\% & $C_{\text{D,1}}$ & $\delta$\% & $C_{\text{L,1}}$ & 
    $\delta$\% & $C_{\text{D,1}}$ & $\delta$\% \\
    \hline\\
    \multirow{10}{*}{$1.2$}&\multirow{5}{*}{0.001} & $20$ & 124672 & 1.8644 & 6.46 & -8.4178 & 0.54 & 1.7101 & 4.33 & -7.9135 & 0.35 \\
    & & $40$ & 149710 & 1.9597 & 1.68 & -8.4576 & 0.07 & 1.7674 & 1.12 & -7.9373 & 0.05 \\
    & & $50$ & 158976 & 1.9740 & 0.96 & -8.4608 & 0.04 & 1.7759 & 0.64 & -7.9392 & 0.02 \\
    & & $80$ & 178960 & 1.9901 & 0.16 & -8.4634 & 0.00 & 1.7866 & 0.11 & -7.9407 & 0.00\\
    & & $100$ & 189702 & 1.9933 & - & -8.4638 & - & 1.7874 & - & -7.9409 & -\\\\
    &\multirow{5}{*}{0.1} & $20$ & 124672 & 1.6215 & 0.23 & -8.3313 & 0.36 & 1.5453 & 0.16 & -7.8553 & 0.22\\
    & & $40$ & 149710 & 1.6188 & 0.06 & -8.3570 & 0.05 & 1.5435 & 0.04 & -7.8703 & 0.03\\
    & & $50$ & 158976 & 1.6183 & 0.04 & -8.3591 & 0.03 & 1.5432 & 0.02 & -7.8715 & 0.02\\
    & & $80$ & 178960 & 1.6179 & 0.01 & -8.3610 & 0.00 & 1.5429 & 0.00 & -7.8726 & 0.00\\
    & & $100$ & 189702 & 1.6178 & - & -8.3614 & - & 1.5429 & - & -7.8728 & -\\
    \hline\\
    \multirow{10}{*}{$9.5$}&\multirow{5}{*}{0.001} & $20$ & 186112 & 0.8483 & 55.99 & -0.0223 & 66.42 & 0.8752 & 50.86 & -0.0396 & 41.27 \\
    & & $40$ & 223210 & 1.5593 & 19.12 & -0.0513 & 22.74 & 1.5107 & 15.18 & -0.0580 & 14.02 \\
    & & $50$ & 236736 & 1.6964 & 12.00 & -0.0577 & 13.15 & 1.6180 & 9.15 & -0.0620 & 8.08 \\
    & & $80$ & 265600 & 1.8814 & 2.41 & -0.0649 & 2.27 & 1.7503 & 1.73 & -0.0665 & 1.39 \\
    & & $90$ & 265600 & 1.9091 & 0.97 & -0.0658 & 0.94 & 1.7690 & 0.68 & -0.0670 & 0.60 \\
    & & $100$ & 280962 & 1.9278 & - & -0.0664 & - & 1.7810 & - & -0.0674 & - \\\\
    &\multirow{5}{*}{0.1} & $20$ & 186112 & 0.2091 & 18.88 & 0.0159 & 42.33 & 0.2600 & 14.79 & -0.0101 & 21.84\\
    & & $40$ & 223210 & 0.2542 & 1.39 & 0.0154 & 37.30 & 0.3026 & 0.83 & -0.0107 & 17.35\\
    & & $50$ & 236736 & 0.2569 & 0.33 & 0.0139 & 24.06 & 0.3045 & 0.20 & -0.0114 & 11.56\\
    & & $80$ & 265600 & 0.2579 & 0.05 & 0.0118 & 4.97 & 0.3052 & 0.03 & -0.0126 & 2.50\\
    & & $90$ & 265600 & 0.2578 & 0.02 & 0.0115 & 2.25 & 0.3053 & 0.05 & -0.0128 & 1.28\\
    & & $100$ & 280962 & 0.2578 & - & 0.0112 & - & 0.3051 & - & -0.0129 & -\\
  \end{tabular}
  \caption{Effect of domain size on drag and lift coefficients for maximum and minimum separation distances (${l^{\ast}} = 1.2$ and $9.5$) and shear Reynolds number ($Re_\gamma = 10^{-3}$ and $10^{-1}$) at $Re_{\text{slip}}=0$. $\delta$ is the percentage error in coefficient, relative to results calculated using the largest domain size (${L^{\ast}}=100$).}
  \label{Domain}
\end{table}
The domain size is first tested to determine a suitable choice of ${L^{\ast}}$ such that the lift and drag forces are negligibly affected by this parameter. We perform a series of simulations where ${L^{\ast}}$ is increased from $20$ to $100$. Simulations are performed for seven selected shear Reynolds numbers; $Re_{\gamma} = (1,2,4,6,8) \times 10^{-3}, 10^{-2}$ and $10^{-1}$; and five selected wall distances; ${l^{\ast}} = 1.2, 2, 4, 6$ and $9.5$; at $Re_{\text{slip}}=0$. The number of mesh points in the domain ($N_{\text{d}}$) is systematically increased with ${L^{\ast}}$ while maintaining a constant progression ratio ($\beta$) and a constant minimum cell thickness in the outer domain. 

Results for the lift and drag coefficients, $C_\text{L,1}$ and $C_\text{D,1}$ respectively, for the minimum and maximum separation distances (${l^{\ast}} = 1.2$ and $9.5$, respectively) and minimum and maximum shear Reynolds numbers ($Re_{\gamma} = 10^{-3}$ and $10^{-1}$) are shown in table \ref{Domain}.  Also shown are the percentage differences ($\delta$) for these coefficients relative to the corresponding values obtained using the maximum domain size (${L^{\ast}}$). We use $\delta$ as an indicator of the coefficient accuracy, noting the limitations of this measure as we approach the maximum domain size.

For $Re_{\gamma} \geq 10^{-2}$, a domain size of ${L^{\ast}} = 50$ is sufficient to capture all the inertial effects responsible for the lift forces, with $\delta$ less than $1\%$ for all separation distances ($1.2 \leq {l^{\ast}} \leq 9.5$) under both non-rotating and freely rotating conditions (additional supporting data not shown). In contrast, for low shear Reynolds numbers, for example the conditions of $Re_{\gamma} = 10^{-3}$, non-rotating (freely-rotating) particles and a domain size of ${L^{\ast}} = 50$, an increase of $\delta$ from $\sim 0.96\%$ $(0.64\%)$ to $\sim 12\%$ $(9.15\%)$ is observed as the separation distance increases from ${l^{\ast}} = 1.2$ to $9.5$. For smaller separation distances (i.e. ${l^{\ast}} = 1.2$), even at low $Re_\gamma$, smaller values for $\delta$ are expected as wall effects dominate outer boundary effects in this near-to-wall regime \citep{Ekanayake}. However, when the distance to the wall is large and $Re_\gamma$ low, the lift force is quite sensitive to the location of the outer boundary.  Noting that the boundary layer thickness around a translating sphere in an unbounded environment is inversely proportional to $\sqrt{Re_\text{slip}}$ \citep{Dandy}, it follows that larger domain sizes are required to minimise outer boundary effects when $Re_\gamma$ is small \citep{Dandy}. Therefore, for the simulations where $Re_{\gamma} < 10^{-2}$, a larger domain size of ${L^{\ast}} = 100$ is used in this study, compared to the domain size of ${L^{\ast}} = 50$ used for the higher $Re_{\gamma}$ situations.  Using this strategy we believe the accuracy of all presented lift coefficients to be better than $1\%$, with the possible exception of values calculated using the combination of the lowest $Re_{\gamma}$ and highest ${l^{\ast}}$ values considered.

For drag, the accuracy of the drag coefficient calculation is largely a function of wall separation distance.  With the selected domain size of ${L^{\ast}} = 50$ that is used for the $Re_{\gamma} \geq 10^{-2}$ results, $\delta < 1\%$ only for ${l^{\ast}} \leq 4$ (additional data not shown). Further away from the wall, a notable dependency on domain size is found, with a $\delta$ variation of $\sim 5\%$ $(2\%)$ to $24\%$ $(12\%)$ for non-rotating (rotating) conditions as ${l^{\ast}}$ increases from $6$ to $9.5$ for $Re_\gamma = 10^{-1}$.  Results for $Re_\gamma = 10^{-3}$ are similar.  Also at these larger separation distances, specifically at ${l^{\ast}}= 8$ and $9.5$, small positive drag coefficients are observed when $Re_{\gamma}=10^{-1}$ (noting that results for ${l^{\ast}} = 4, 6, 8$ are not presented in the table). The small positive drag coefficients and larger $\delta$ values found at large separation distances are caused by small errors in $F_\text{D,1}^{\ast}$ which are amplified when expressed as a relative drag coefficient error, because as the separation distance increases, both the coefficients and the force approach zero.  Note however that while the drag coefficient errors are higher at the largest separation distances, an $11\%$ change in domain size from ${L^{\ast}}=90$ to $100$ results in less than a $1\%$ change in drag coefficient (for $Re_{\gamma}=10^{-3}$), suggesting that the results are close to independent of domain size. In summary, while the errors in the calculated drag coefficients are larger than for lift, the largest errors occur under conditions in which the drag force is low.  Under conditions in which the drag force is appreciable, the error in the drag force coefficient is similar to that reported for the lift force coefficient.

\subsection{Mesh Dependency}\label{sec:Mesh Dependency}

\begin{figure}
\centering
\includegraphics[width=0.8\columnwidth]{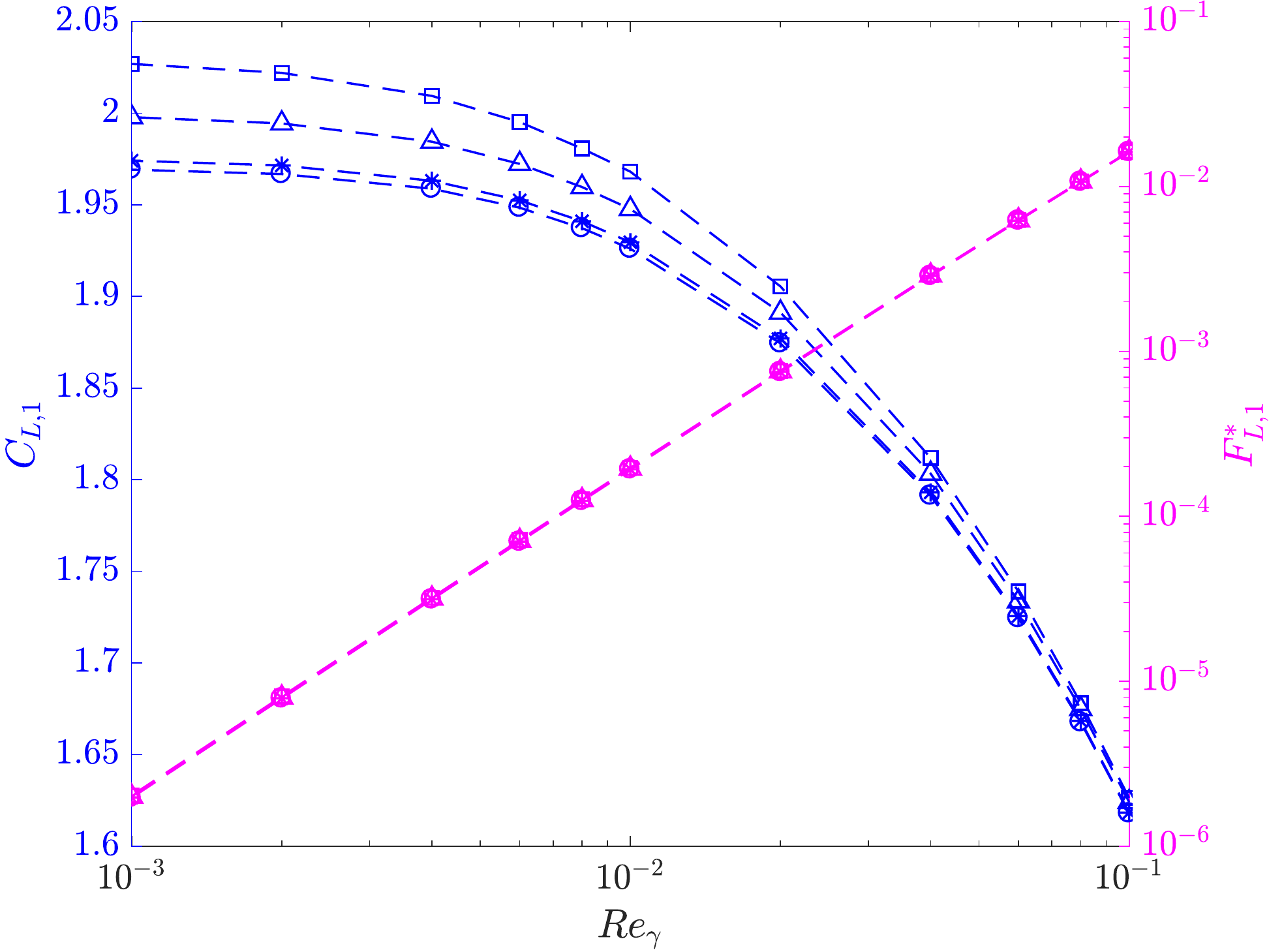}
\caption{Effect of mesh resolution around the sphere on $C_\text{L,1}$ for a non-rotating particle at ${l^{\ast}}=1.2$. $N_\text{p},N_\text{d}$ = 15 (\protect\hollowsquaresolidline); 20 (\protect\hollowtrainglesolidline); 25 (\protect\hollowstarsolidline); 30 (\protect\hollowcirclesolidline).} 
\label{Mesh1}
\end{figure}

The effect of mesh resolution within the boundary layers surrounding the sphere is examined in this section. The number of inflation layers around the sphere and the number of cells on the sphere surface were adjusted by varying $N_{\text{p}}$ while maintaining the same progression rate $\alpha$ in the bounding box. Concurrently, $N_{\text{d}}$ was also changed, maintaining $N_{\text{d}} = N_{\text{p}}$ for all cases.

Figure \ref{Mesh1} shows the resulting lift coefficients and non-dimensionalised lift forces for four mesh refinement levels around the sphere and for the conditions of ${l^{\ast}}=1.2$, a domain size of ${L^{\ast}}=50$, and a range of shear Reynolds numbers. As $F^{\ast}_{L,1}$ reduces to zero, $C_\text{L,1}$ asymptotes to different finite values as $Re_{\gamma}$ approaches zero. A considerable variation of $C_\text{L,1}$ with mesh refinement is observed, particularly at low $Re_{\gamma}$, however the relative change in $C_\text{L,1}$ decreases as $N_\text{p}$ and $N_\text{d}$ increase. Indeed, increasing  $N_\text{p}$ and $N_\text{d}$ from $25$ to $30$ only changes $C_\text{L,1}$ by a small amount, but concurrently increases the total cell count $N_\text{t}$ from 158,976 to 310,500, significantly increasing computational memory requirements ($\sim 1.2$ Tb).  Noting that for low $Re_{\gamma}$ the force will be negligible anyway, and that across the entire $Re_{\gamma}$ range the difference between the $N_\text{p}, N_\text{d}=25$ and $N_\text{p}, N_\text{d}=30$ results is small anyway, we employed the mesh with $N_\text{p},N_\text{d} = 25$ for the remainder of the study.

\section{Results and Discussion}\label{sec:Results and Discussion}

In this section we first present results for the lift force (Section \ref{sec:Lift force}), then for the drag force (Section \ref{sec:Drag force}), before finally using both results to assess the importance of the zero-slip coefficients in determining the slip velocity and lift force acting on particles translating near walls under linear shear conditions (Section \ref{sec:Application}).

\subsection{Lift force}\label{sec:Lift force}

\begin{figure}
\begin{subfigure}{\linewidth}
\centerline{\includegraphics[width=0.80\linewidth]{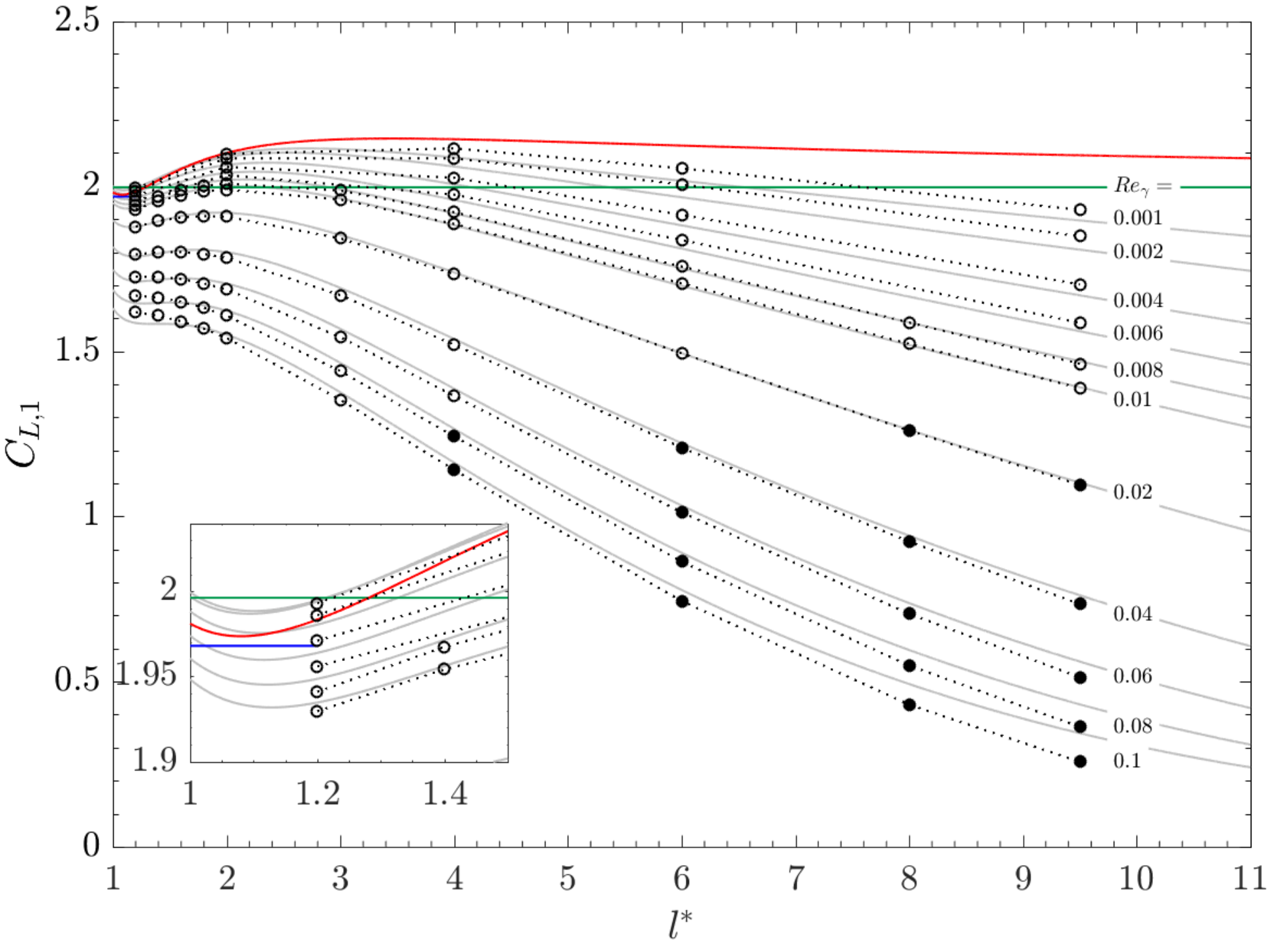}}
\caption{}
\label{Lift_slip0_omega0_fit}
\end{subfigure}
\begin{subfigure}{\linewidth}
\centerline{\includegraphics[width=0.80\linewidth]{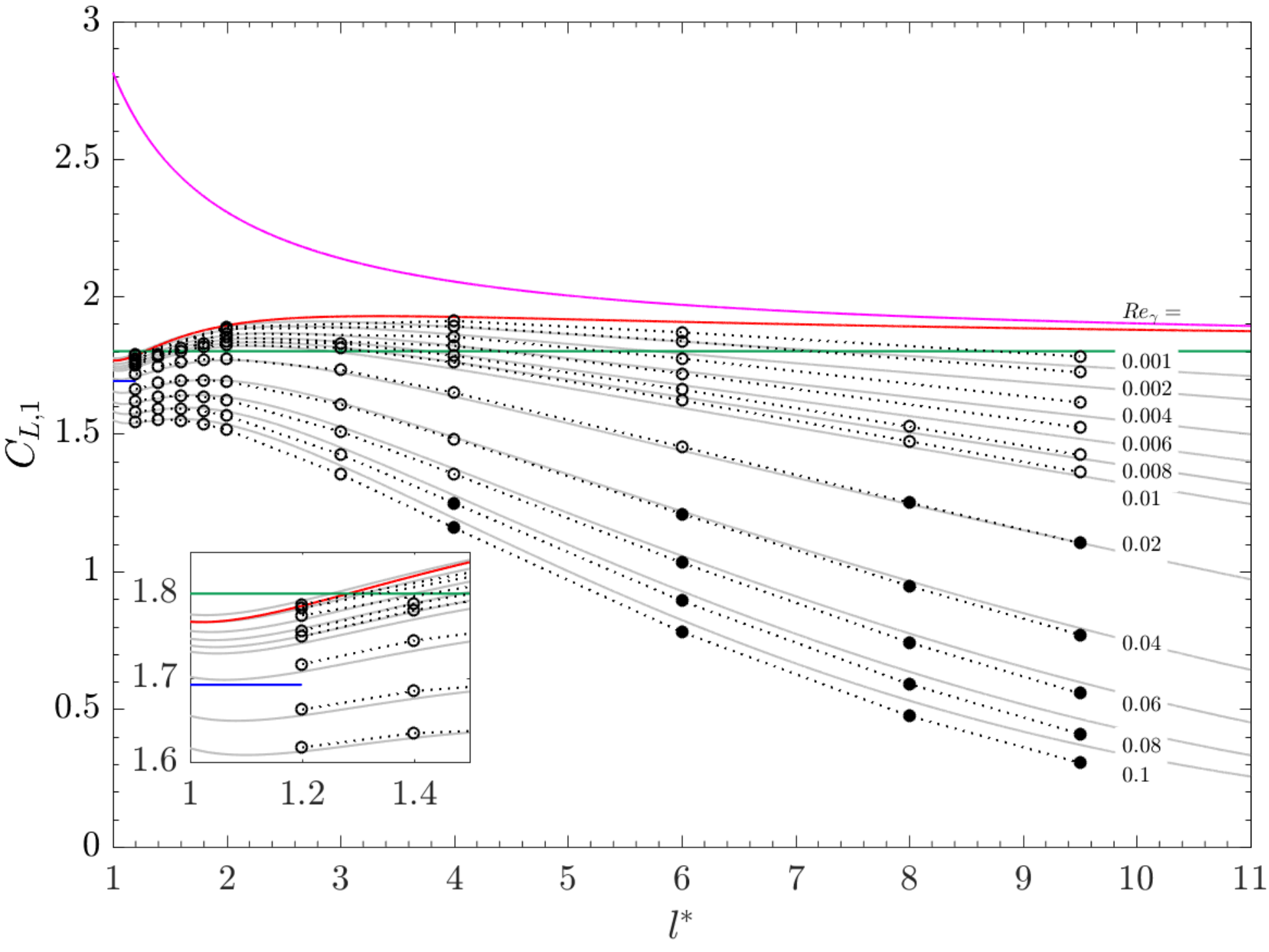}}
\caption{}
\label{Lift_slip0_torque0_fit}
\end{subfigure}
\caption{Lift coefficient ($C_{\text{L,1}}$) for different shear Reynolds number as a function of non-dimensional separation distance (${l^{\ast}}$) for (a) non-rotating and (b) freely-rotating spheres. Simulations: Inner Region (\protect\hollowcircleline), Outer Region (\protect\solidcircleline). Analytical Eq. (\ref{Cox1}), (\ref{Cox2}) (\protect\greenline) from \citet{CoxHsu}, Eq. (\ref{CM1}), (\ref{CM2}) (\protect\redline) from \citet{CherukatMcLaughlin}, Eq. (\ref{KL1}), (\ref{KL2}) (\protect\blueline) from \citet{Krishnan}, Eq. (\ref{MD1}) (\protect\magentaline) from \citet{Magnaudet1}. Present numerical Fit: Eq. (\ref{0slipnewmodel}) (\protect\Grayline). Present correlation results for $Re_\gamma=0$ are coincident with the \citet{CherukatMcLaughlin} results and not shown on this figure.}
\label{Lift_slip0_fit}
\end{figure}

\begin{figure}
\begin{subfigure}{\linewidth}
\centerline{\includegraphics[width=0.8\columnwidth]{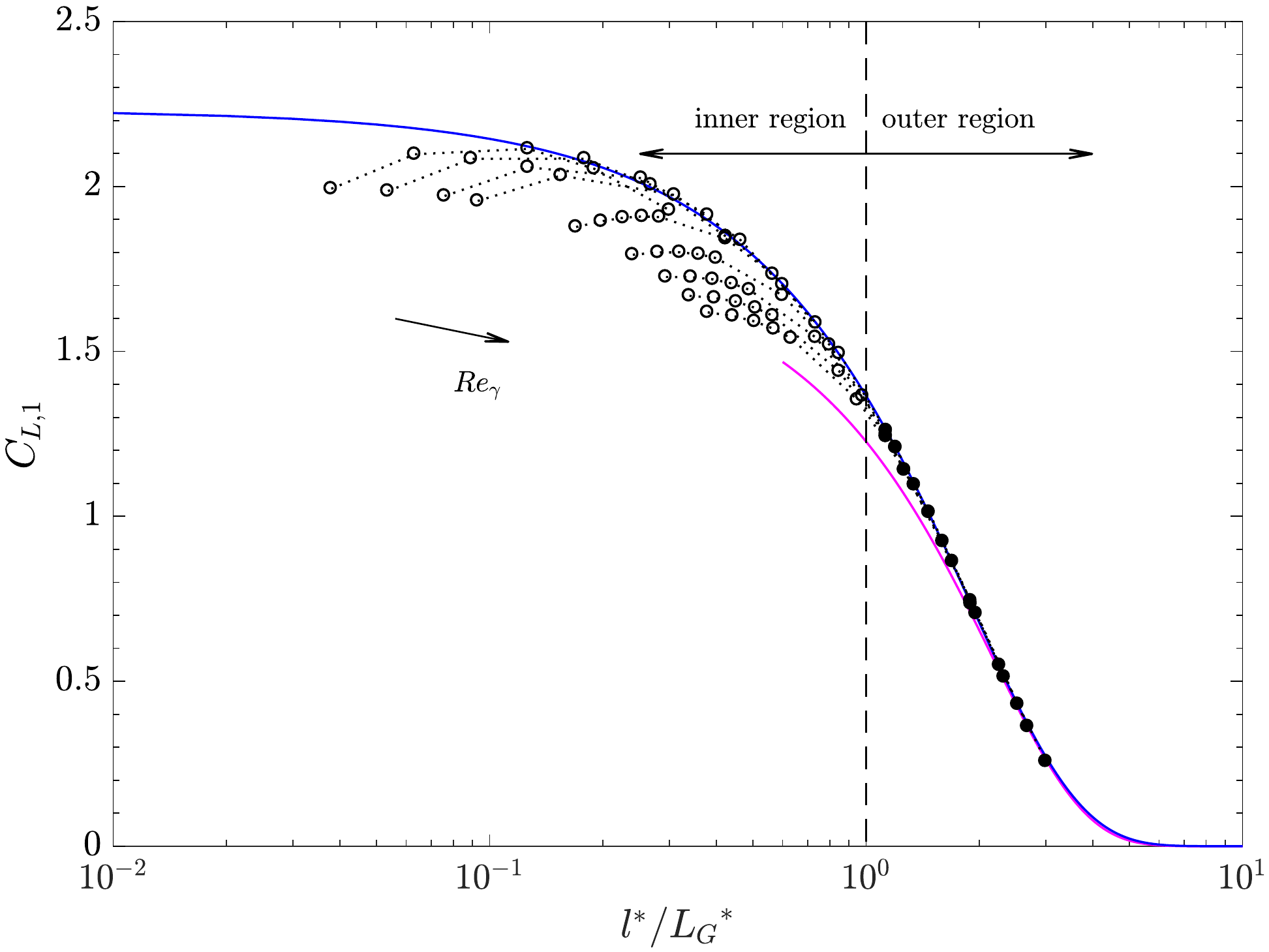}}
\caption{}
\label{Lift_slip0_omega0_saffman}
\end{subfigure}
\begin{subfigure}{\linewidth}
\centerline{\includegraphics[width=0.8\columnwidth]{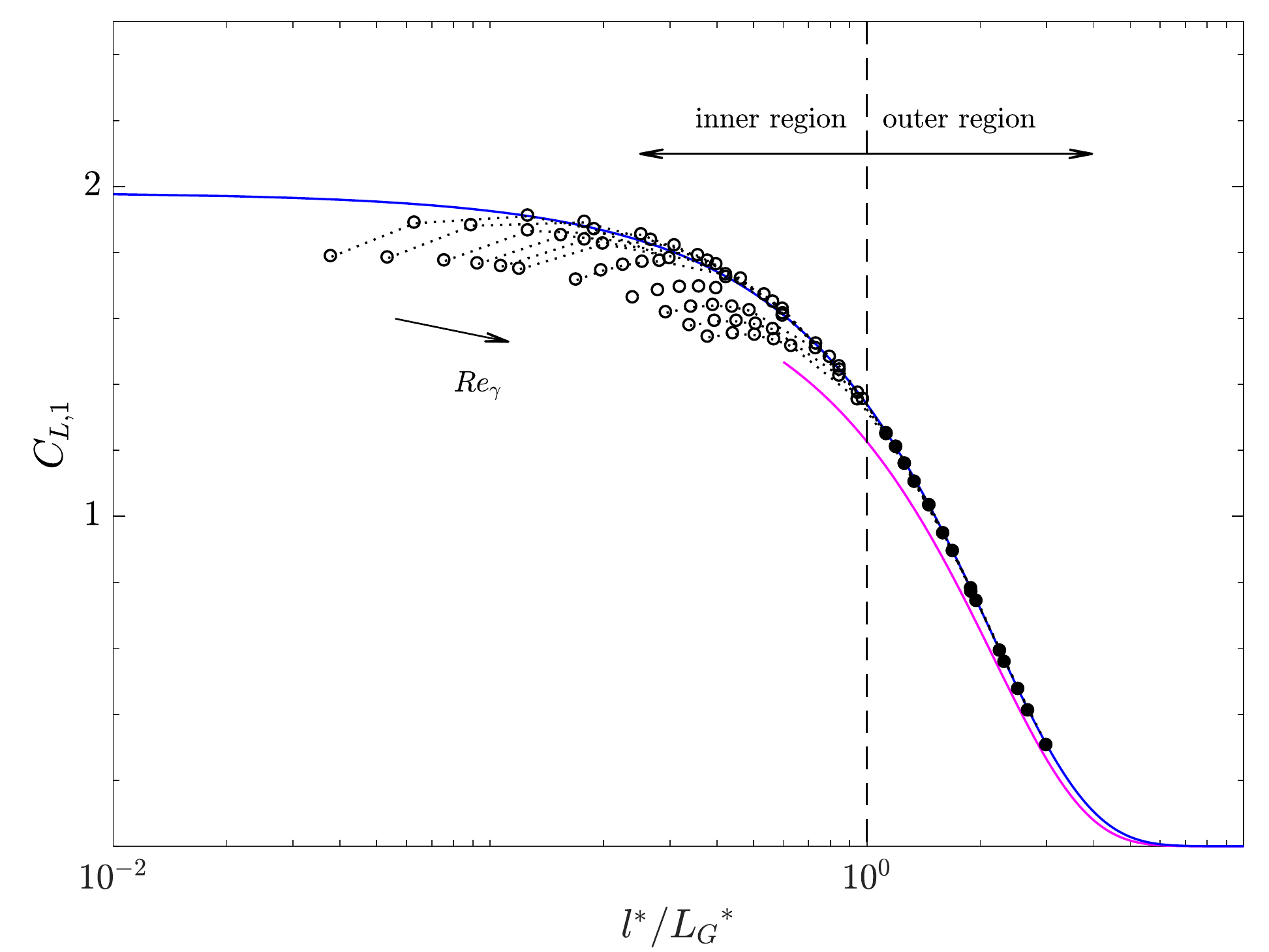}}
\caption{}
\label{Lift_slip0_torque0_saffman}
\end{subfigure}
\caption{Lift coefficient ($C_{\text{L,1}}$) for shear Reynolds number values of 1, 2, 4, 6, 8, 10, 20, 40, 60, 80 and 100; $\times 10^{-3}$ (the arrow
indicates the direction of increasing $Re_\gamma$) as a function of separation distance non-dimensionalised by Saffman length scale ($l^{\ast}/{L_{\text{G}}}^{\ast}$) for a (a) non-rotating and (b) freely-rotating particle.  Simulations: inner region (\protect\hollowcircleline), outer region (\protect\solidcircleline). Present numerical Fit: Eq. (\ref{outerregion}) (\protect\blueline). \citet{Asmolov99} results for outer region (\protect\magentaline).}
\label{Lift_slip0_saffman}
\end{figure}

In figure \ref{Lift_slip0_fit}, lift coefficients $C_\text{L,1}$ computed for both non-rotating and a freely rotating particles are plotted as a function of separation distance (${l^{\ast}}$). The numerical results are compared with the available inner region correlations listed in table \ref{InnerLiftSummary} that are valid for $Re_\gamma \ll 1$ and $Re_{\text{slip}} = 0$. For $Re_{\gamma} < 10^{-2}$, the numerically computed lift forces in the region close to the wall (${l^{\ast}} < 2$) agree reasonably well with the asymptotic values predicted by the analytical solutions derived for low Reynolds numbers \citep{CherukatMcLaughlin,Krishnan}. Specifically, the lowest shear Reynolds number simulation conducted at the smallest distance to the wall (${l^{\ast}} = 1.2$) gives a $C_{\text{L,1}}$ of $1.993$ ($1.787$) for a non-rotating (freely-rotating) particle, which is only $\sim 1.3\%$ ($5.2\%$) higher than the asymptotic value of 1.9680 (1.6988) predicted for a non-rotating (freely-rotating) particle at ${l^{\ast}} = 1$ \citep{Krishnan}. The computed lift coefficients also agree very well with the low Reynolds number theoretical values of 1.9834 (1.7854) for non-rotating (freely-rotating) particles at ${l^{\ast}} = 1.2$, as predicted by \citet{CherukatMcLaughlin}. However, as illustrated in figure \ref{Lift_slip0_torque0_fit}, for a freely rotating particle the \citet{Magnaudet1} lift correlation predicts a larger coefficient of 2.81 at ${l^{\ast}} = 1$ which is inconsistent with the numerical data and the available theories in this region. As explained in \citet{Shi2}, this over-prediction of the lift coefficient near the wall is due to the neglect of the higher order separation distance terms ($\mathcal{O}(1/{l^{\ast}}) > 2$) that are significant when representing lift  in the vicinity of the wall. At large $l^{\ast}$ the \citet{Magnaudet1} lift correlation gives $C_{\text{L,1}}=1.8$, a result that is consistent with the large ${l^{\ast}}$ value from the \citet{CoxHsu} lift correlation.  However, like all inner-region based models, this large separation limit only has practical relevance for very small Reynolds numbers.

\begin{table}
    \centering
    \begin{tabular}{ccc}
    Coefficient & Non-rotating & Freely-rotating\\
    $\lambda_1$ & 0.9300 & 0.9250\\
    $\lambda_2$ & -0.8800 & -0.3500\\
    $\lambda_3$ & -0.0300 & -0.0135\\
    $\lambda_4$ & -1000 & -7000\\
    $\lambda_5$ & 2.231 & 1.982\\
    $\lambda_6$ & -0.1054 & -0.1150\\
    $\lambda_7$ & -0.3859 & -0.2771\\
    ${\lambda_8}^\text{\textdagger}$ & 1.0575 & 0.89934\\
    ${\lambda_9}^\text{\textdagger}$ & -2.4007 & -1.9610\\
    ${\lambda_{10}}^\text{\textdagger}$ & 1.3174 & 1.0216\\
  \end{tabular}
  \caption{Coefficients for Eq. (\ref{0slipnewmodel}) for a non-rotating and freely-rotating particle. $^\text{\textdagger}$coefficients from \citet{CherukatMcLaughlin} \& \citet{CherukatMcLaughlinCorrection}.}
  \label{lift_coefficients}
\end{table}

As $Re_{\gamma}$ increases and inertial effects become more significant, the computed lift coefficients deviate significantly from the available theoretical values. Even when the particle is far away from the wall but still within the inner region (eg. results for $Re_\gamma < 10^{-2}$), the computed lift force coefficient decreases with shear rate, in contrast to the available inner region models which give coefficients that are independent of $Re_\gamma$. For example, the \citet{CherukatMcLaughlin} model gives a $C_{\text{L,1}}$ of 2.0069 (1.8065) for a non-rotating (freely rotating) particle as ${l^{\ast}} \rightarrow \infty$. Eventually, with increasing shear rate and increasing separation distance, the walls move to the outer region ($l^{\ast}/{L_\text{G}}^{\ast} > 1$) and unsurprisingly, the inner region based models do not capture the lift coefficient variation well at all. Hence the available inner region based theoretical lift models overestimate the lift coefficient for all but the lowest shear Reynolds numbers, for particles both near and (especially) further away from the wall.

The discrepancy between simulation and theory arises for the inner region based models because they use a first-order expansion for the disturbance velocity that cannot satisfy the boundary conditions at infinity \citep{CherukatMcLaughlin}. For many applications relevant to cell sorting in microfluidics, or to understand mechanical phenomena like particle deposition and fouling at $Re_\gamma \lesssim \mathcal{O}(1)$, this is problematic, and a single equation for the wall-induced lift force valid across both the inner and outer regions would be beneficial. Therefore, as well as providing a shear-based inertial correction for the analytical inner region model of \citet{CherukatMcLaughlin}, we here propose a model that simultaneously captures the outer region behaviour.

In this vein, we first plot the computed lift force coefficients against the separation distance, now normalised using Saffman's length scale ($L_{\text{G}}^{\ast}$). Recall that this length scale defines the boundary between the inner and outer regions in this problem. These results, shown in figure \ref{Lift_slip0_saffman}, indicate that the outer region data ($l^{\ast}/L_{\text{G}}^{\ast} > 1$) collapses to a single curve for the selected range of shear Reynolds numbers, similar to the results given for a freely rotating particle at $Re_\gamma \ll 1$ presented by \citet{Asmolov99}. The difference between a non-rotating and freely rotating lift coefficient value is less significant in the outer region, particularly at $l^{\ast}/L_{\text{G}}^{\ast} \gg 1$.
Utilising this outer region behaviour, together with the existing low $Re_\gamma$ inner region \citet{CherukatMcLaughlin} results, the following lift correlation is proposed for particles moving near a wall under zero slip conditions:
\begin{equation}
    {C_\text{L,1}} = f_\text{1}(Re_\gamma){C_\text{outer}}(l^{\ast}/{L_\text{G}}^{\ast}) + f_\text{2}(Re_\gamma){C_\text{inner}}({1/l^{\ast}})
    \label{0slipnewmodel}
\end{equation}
where
\begin{gather}
    f_\text{1}(Re_\gamma) = {\lambda_{1}}\exp{({\lambda_{2}} Re_\gamma)}+{\lambda_3}\exp{({\lambda_4} Re_\gamma)};\\
    \label{outerregion}
    {C_\text{outer}}(l^{\ast}/L_\text{G}^{\ast}) = {\lambda_5}\exp\big[{{\lambda_6}\big({l^{\ast}}/{L_\text{G}^{\ast}}\big)^2+{\lambda_7}\big({l^{\ast}}/{L_\text{G}^{\ast}}\big)}\big];\\
    f_\text{2}(Re_\gamma) =1+\sqrt{Re_\gamma};\\
    {C_\text{inner}}({1/l^{\ast}})={\lambda_8}(1/{l^{\ast}})+{\lambda_9}(1/{l^{\ast}})^{2}+{\lambda_{10}}(1/{l^{\ast}})^{3}.
\end{gather}
Coefficients for the above functions are listed in table \ref{lift_coefficients} for both non-rotating and freely rotating particles.

The functions used in equation (\ref{0slipnewmodel}) represent different limits. Within the inner region, the first term of equation (\ref{0slipnewmodel}) rapidly reduces to zero as both $Re_\gamma \rightarrow 0 $ and $l^{\ast} \rightarrow 1$, leaving the second part of the equation to be dominant. The function ${C_\text{inner}}$ consists of the non-zero degree polynomial terms of the \citet{CherukatMcLaughlin} model. Here $f_\text{2}(Re_\gamma)$, which is independent of wall distance, captures inertial effects when the particle is close to the wall. At ${l^{\ast}} = 1$ and $Re_{\gamma} = 0$, the product of $f_\text{2}(Re_\gamma){C_\text{inner}}$ reduces to the constant value of $2.0069$ ($1.8065$), consistent with the \citet{CherukatMcLaughlin} result for a non-rotating (freely-rotating) particle. With increasing shear rate and wall distance, the first term of the Eq. (\ref{0slipnewmodel}) becomes dominant and captures the outer region variation. This term gives zero $C_{\text{L,1}}$ as ${l^{\ast}} \rightarrow \infty$.

Figure \ref{Lift_slip0_fit} shows that the proposed model captures the simulated lift coefficients accurately over most of the $Re_\gamma$ range and separation distances considered in this study. For low shear rates ($Re_\gamma \sim 10^{-3}$), the model performs well, slightly underestimating the simulated results with a maximum deviation of $1.6\%$ $(3.2\%)$ for a non-rotating (freely-rotating) particle at distances far from the wall. For high shear rates ($Re_\gamma \sim 10^{-1}$), the model generally overestimates the simulated results, with a maximum deviation of $41\%$ $(21\%)$ for a non-rotating particle (freely-rotating particle) occurring when the particle is furthest from the wall, that is at $l^{\ast} = 9.5$. This deviation rapidly reduces down to 4.5\% ($5.6\%$) for a non-rotating (freely-rotating) particle when the separation distance decreases to $l^{\ast} = 6$, showing that even for the relatively high $Re_\gamma = 10^{-1}$, the proposed model still accurately captures the lift force variation provided that the particle is within a few diameters of the wall. Note that as discussed in Section \ref{sec:Domain Size Dependency}, we expect the numerical data to have the highest errors under the same high $Re_\gamma$ and $l^{\ast}$ conditions that generate the largest deviations between correlation and simulation - conditions that also result in low lift forces.  Hence, the proposed model captures the variation of lift coefficient with $Re_\gamma$ and separation distance reasonably well across all conditions considered, but is most accurate when the resulting lift force is most significant.

\subsection{Drag force}\label{sec:Drag force}

\begin{figure}
\begin{subfigure}{\linewidth}
\centerline{\includegraphics[width=0.85\textwidth]{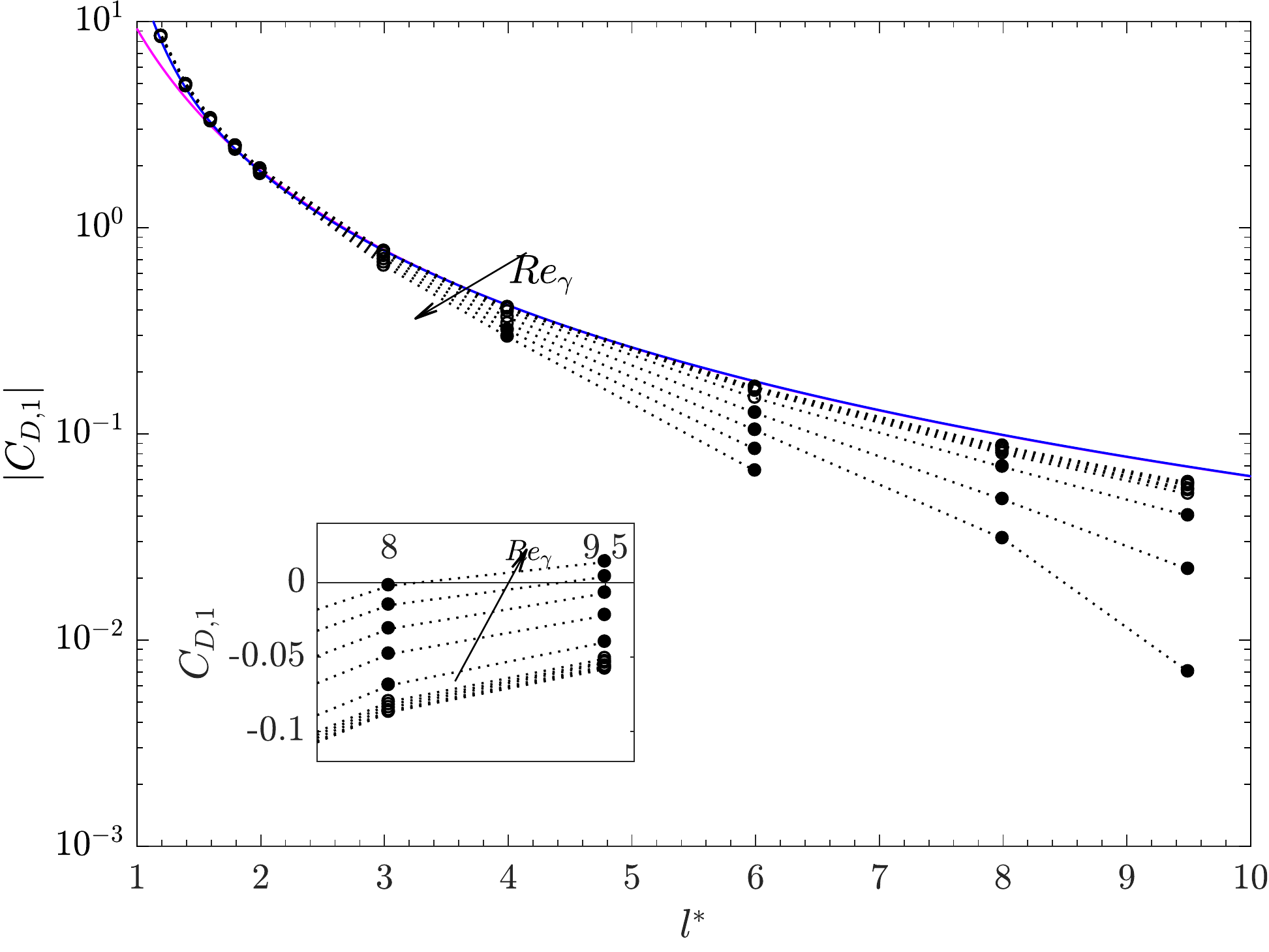}}
\caption{}
\label{Dragcd_slip0omega0}
\end{subfigure}
\begin{subfigure}{\linewidth}
\centerline{\includegraphics[width=0.85\textwidth]{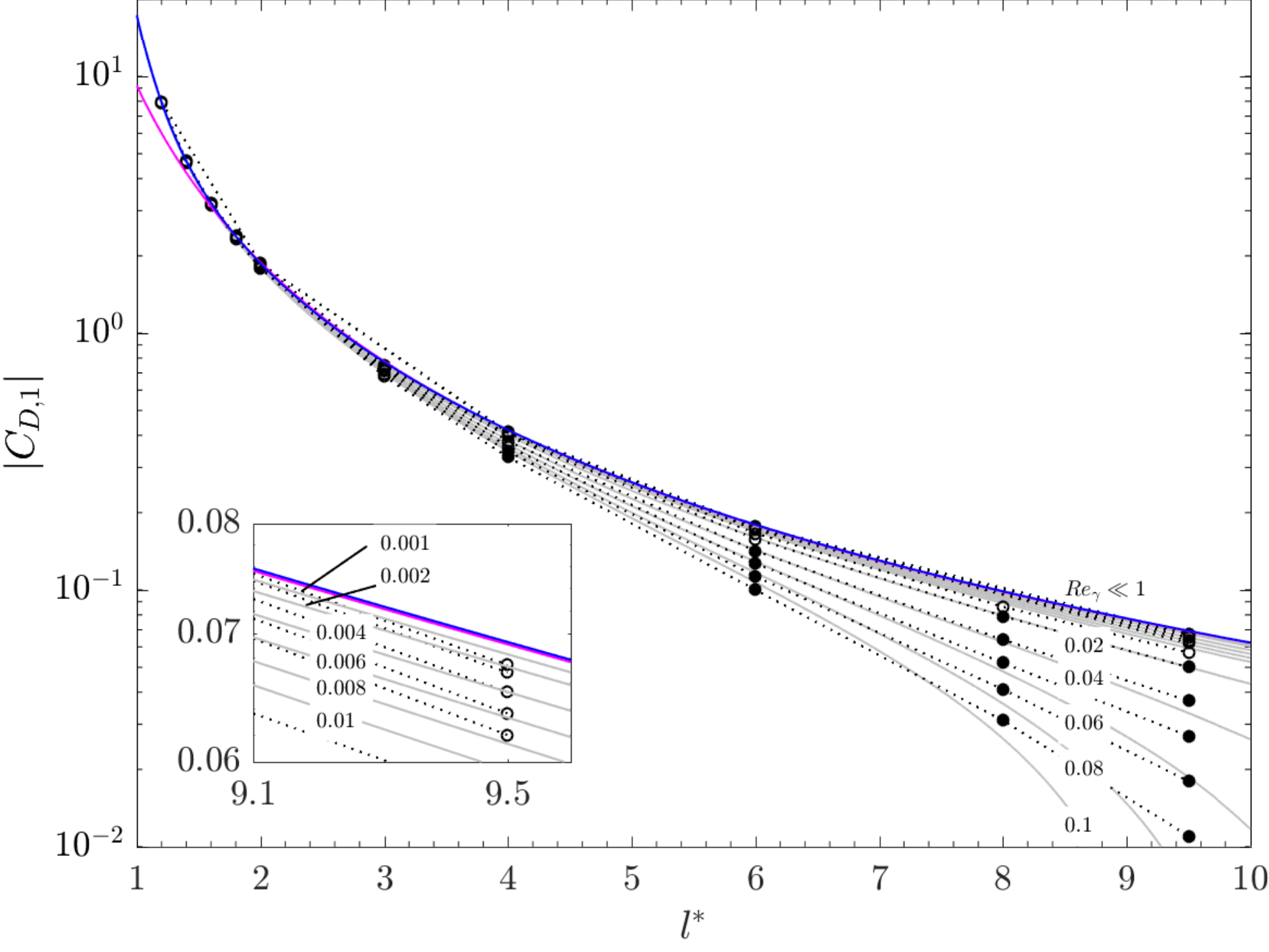}}
\caption{}
\label{Dragcd_slip0torque0}
\end{subfigure}
\caption{Magnitude of drag force coefficient in the absence of slip ($|C_{\text{D,1}}|$) (a) for a non-rotating (b) freely rotating particle. The arrow
indicates the direction of increasing $Re_\gamma$. Simulations: Inner Region (\protect\hollowcircleline), Outer Region (\protect\solidcircleline). Analytical prediction of \citet{Magnaudet1} Eq. (\ref{eq:dragCD1}) (\protect\magentaline). Present numerical fits for $Re_\gamma \ll 1$: Eq. (\ref{eq:dragCD2}) (\protect\blueline) and finite $Re_\gamma$ Eq. (\ref{eq:dragCD3}) (\protect\Grayline).}
\label{Dragcd_slip0}
\end{figure}

The drag force on a spherical particle moving parallel to a wall and in the absence of slip is considered in this section. Under these conditions the presence of the wall (combined with the shear rate) produces a force on the particle that is in the opposite direction to the flow, and which rapidly decays as the separation distance between the sphere and wall increases. Note that this wall-shear drag causes otherwise force-free, neutrally-buoyant particles to lag the flow when moving in close proximity to a wall. 

Figure \ref{Dragcd_slip0} shows the variation in drag coefficient $C_\text{D,1}$ for shear Reynolds numbers in the range $Re_{\gamma} = 10^{-3} - 10^{-1}$ as a function of wall separation distance. Results for both non-rotating and freely-rotating conditions are shown. 

Figure \ref{Dragcd_slip0omega0} is specific to non-rotating particles. The highest drag coefficient magnitudes are observed close to the wall, with a rapid decrease occurring with increasing wall distance or increasing $Re_\gamma$. For the largest separation considered (${l^{\ast}} = 9.5$), small positive $C_\text{D,1}$ values of $\mathcal{O}(10^{-3})$ - $\mathcal{O}(10^{-2})$ are reported for $Re_\gamma=10^{-1}, 8 \times 10^{-2}$ as illustrated in the inset of figure \ref{Dragcd_slip0omega0}. As discussed in Section \ref{sec:Domain Size Dependency}, these small positive drag coefficients are more likely to be due to the limited domain size employed in the simulations (numerical accuracy), rather than being physically meaningful.

Figure \ref{Dragcd_slip0torque0} shows the drag coefficient for a freely rotating particle as a function of wall distance. The highest negative values for $C_\text{D,1}$ are obtained when the particle is close to the wall, reducing to zero with increasing wall distance, in a manner similar to that observed for a non-rotating particle. The highest reported $C_\text{D,1}$ values are $\sim 6\%$ less than the values reported for a non-rotating particle. The simulated drag coefficient results are also compared with the inner region analytical drag correlation of \citet{Magnaudet1} for a freely-rotating particle at low $Re_{\gamma} \ll 1$. The computed drag coefficients are in good agreement with the correlation values for Reynolds numbers $Re_{\gamma} < 10^{-2}$ and for wall distances ${l^{\ast}>1.6}$. However, as the separation distance decreases (${l^{\ast}} \rightarrow 1$), the theoretical predictions of Eq. (\ref{eq:dragCD1}) underestimate the drag, resulting in a difference of $\sim32\%$ at ${l^{\ast}} = 1.2$. This could again be due to the neglect of higher order terms in the analytical solution, Eq. (\ref{eq:dragCD1}). In support of this, in the same study by \citet{Magnaudet1}, but for zero shear flows ($Re_\gamma = 0$), the effect of higher-order ${l^{\ast}}$ terms on $C_\text{D,2}$ was examined in the context of small separation distances (${l^{\ast}} < 3$), and it was found that a $\sim60\%$ difference in this drag coefficient occurred when using $\mathcal{O}(({1/l^{\ast}})^3)$ over $\mathcal{O}(({1/l^{\ast}})^5)$ terms in their analysis, for ${l^{\ast}} \rightarrow 1$. Similar higher order terms for the $C_\text{D,1}$ drag coefficient examined in our study are not available in the literature, however.

The freely rotating drag coefficients shown in Figure \ref{Dragcd_slip0torque0} also display a dependence on shear rate, as well as separation distance. Generally studies of freely translating particles and fixed particles give inertial corrections for $C_\text{D,1}$ and $C_\text{D,2}$ in terms of the slip velocity ($Re_\text{slip}$) \citep{Magnaudet1, Kurose99}. However, in the present zero-slip context these inertial corrections are not relevant, and instead an inertial correction based on $Re_{\gamma}$ is required.

Hence, to accurately predict the variation of $C_\text{D,1}$ over the shear rates and separation distances considered in this study, we propose two modifications to existing theories.

Firstly, to capture the drag force variation close to the wall at $Re_\gamma \ll 1$, Eq. (\ref{eq:dragCD1}) is modified as:
\begin{equation}
 C_\text{D,1}^{'}=-\dfrac{15\pi}{8}\bigg(\dfrac{1}{l^{\ast}}\bigg)^2\bigg[1+\dfrac{9}{16}\bigg(\dfrac{1}{l^{\ast}}\bigg)+0.5801\bigg(\dfrac{1}{l^{\ast}}\bigg)^2 -3.34\bigg(\dfrac{1}{l^{\ast}}\bigg)^3+4.15\bigg(\dfrac{1}{l^{\ast}}\bigg)^4\bigg]
\label{eq:dragCD2}
\end{equation}
where $C_\text{D,1}^{'}$ consists of the terms given by \citet{Magnaudet1} in Eq. (\ref{eq:dragCD1}), combined with three new higher-order terms in separation distance derived through a numerical fitting to our simulation results for $Re_{\gamma} \ll 1$. The proposed correlation accurately captures the computed drag coefficient down to a minimum separation distance of ${l^{\ast}=1.2}$. When the particle is in contact with the wall ($l^{\ast}=1$), the current model predicts a finite drag coefficient $C_\text{D,1}$ of -17.392 (compared to a value of -9.204 from Eq. (\ref{eq:dragCD1})). Whilst the proposed model performs better than the \citet{Magnaudet1} correlation at $l^{\ast}=1.2$, for the limiting case of $l^{\ast}=1$ there is no asymptotic value available for $C_\text{D,1}$ upon which to validate the new expression.

Secondly, we present an inertial correction to the proposed low $Re_{\gamma}$ equation (Eq. \ref{eq:dragCD1}), based on fits to our data, that predicts the variation of $C_\text{D,1}$ with finite $Re_{\gamma}$. The final wall-shear drag correlation can be written:
\begin{equation}
C_\text{D,1}=C_\text{D,1}^{'}-(3.001Re_\gamma^2 -1.025Re_\gamma)
\label{eq:dragCD3}
\end{equation}
where $C_\text{D,1}^{'}$ is given via equation (\ref{eq:dragCD1}). Although, the new inertial correction accurately captures the shear dependence of $C_\text{D,1}$ across all considered separations at low $Re_\gamma$, far away from from the wall ($l^{\ast} > 8$) the coefficient is underestimated for $Re_\gamma > 6 \times 10^{-2}$. Note however that these are the same conditions that result in the poorest accuracy of the computed  $C_\text{D,1}$, as previously discussed.

\subsection{Application of the correlations \label{sec:Application}}

In this section, we analyse the movement of a freely translating, neutrally-buoyant particle in a linear shear flow using the proposed force correlations.  This is accomplished in two parts.  First, we use the new drag correlation to find the slip velocity of a particle moving near a wall.  This is then combined with the new lift correlation to find the lift force acting on the particle and migration velocity of the particle, accounting for both shear and slip effects.

\subsubsection{Slip velocity}

\begin{figure}
\begin{subfigure}{\linewidth}
\centerline{\includegraphics[width=0.55\textwidth]{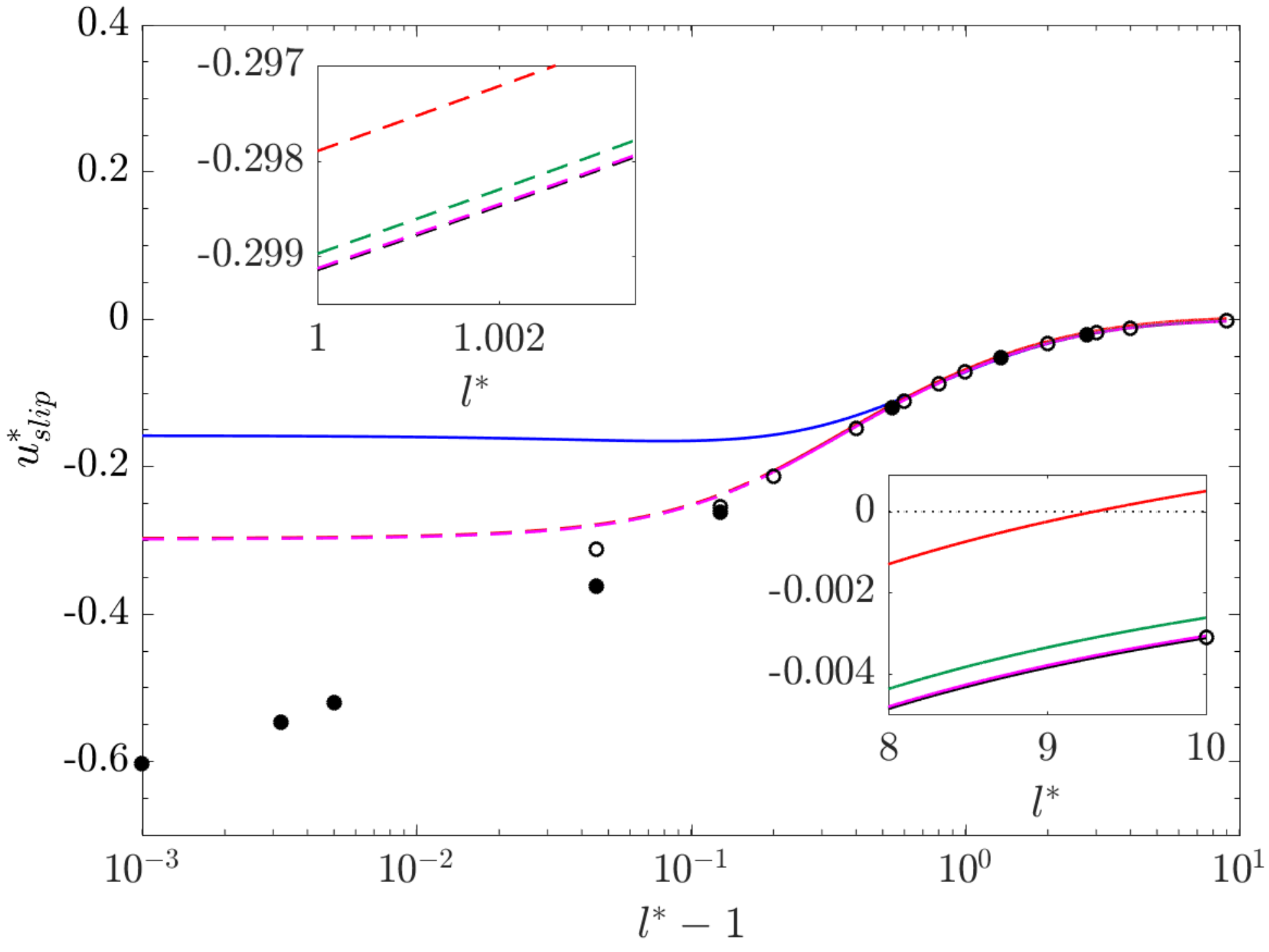}}
\caption{}
\label{sliponly}
\end{subfigure}
\centering
\begin{subfigure}{0.31\linewidth}
\centerline{\includegraphics[width=\textwidth]{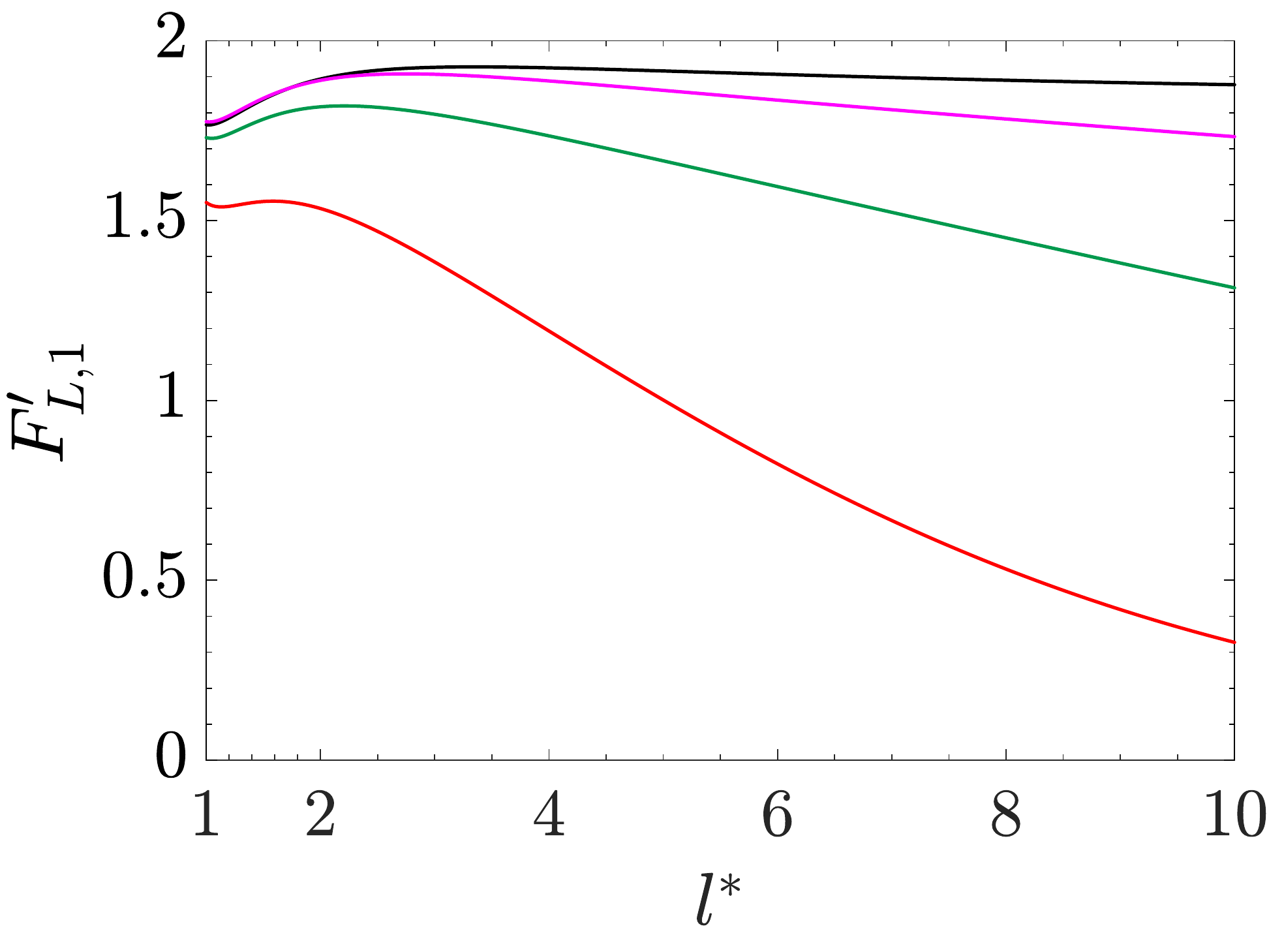}}
\caption{}
\end{subfigure}
\begin{subfigure}{0.31\linewidth}
\centerline{\includegraphics[width=\textwidth]{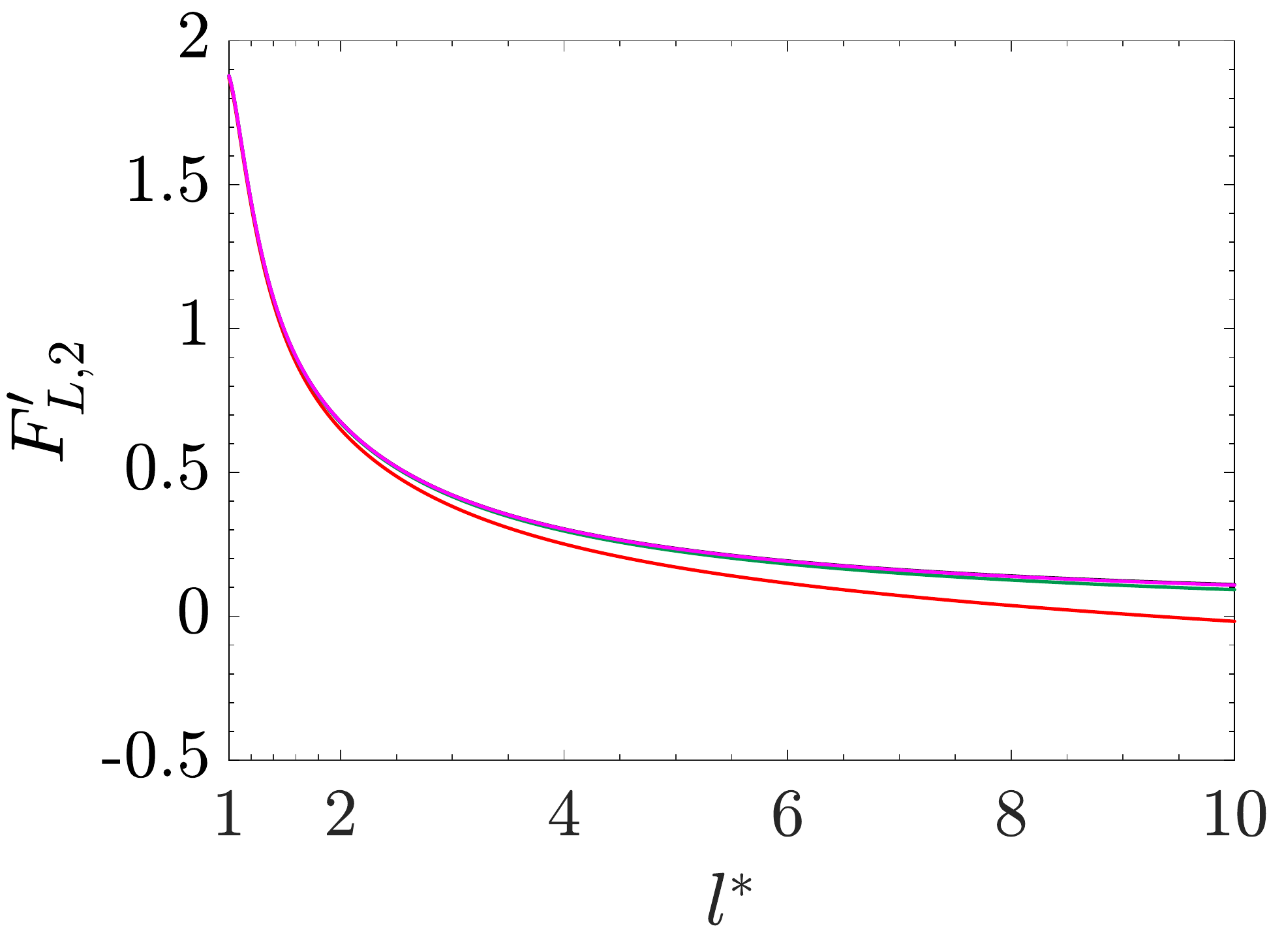}}
\caption{}
\end{subfigure}
\begin{subfigure}{0.31\linewidth}
\centerline{\includegraphics[width=\textwidth]{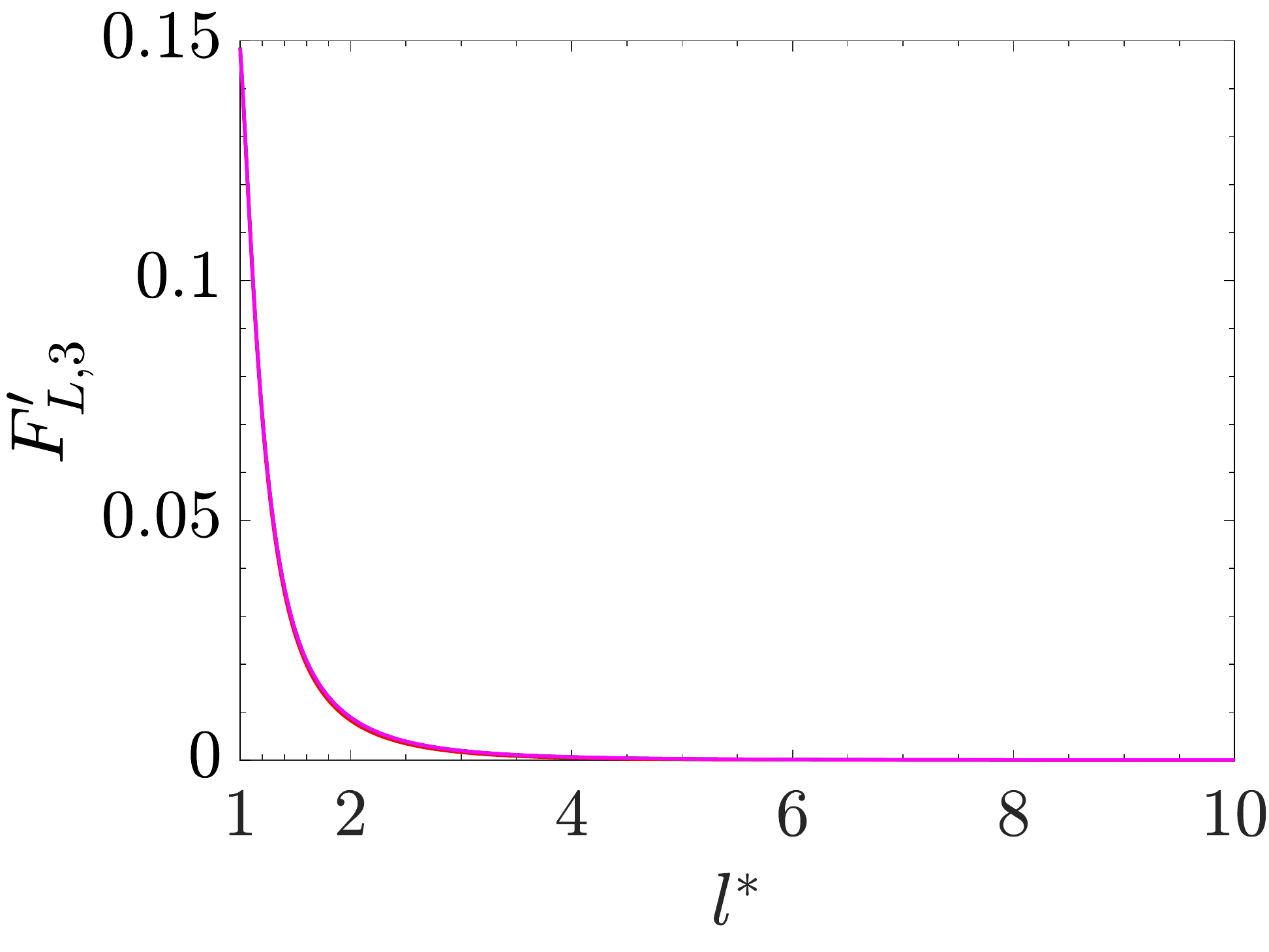}}
\caption{}
\end{subfigure}
\begin{subfigure}{0.55\linewidth}
\centerline{\includegraphics[width=\textwidth]{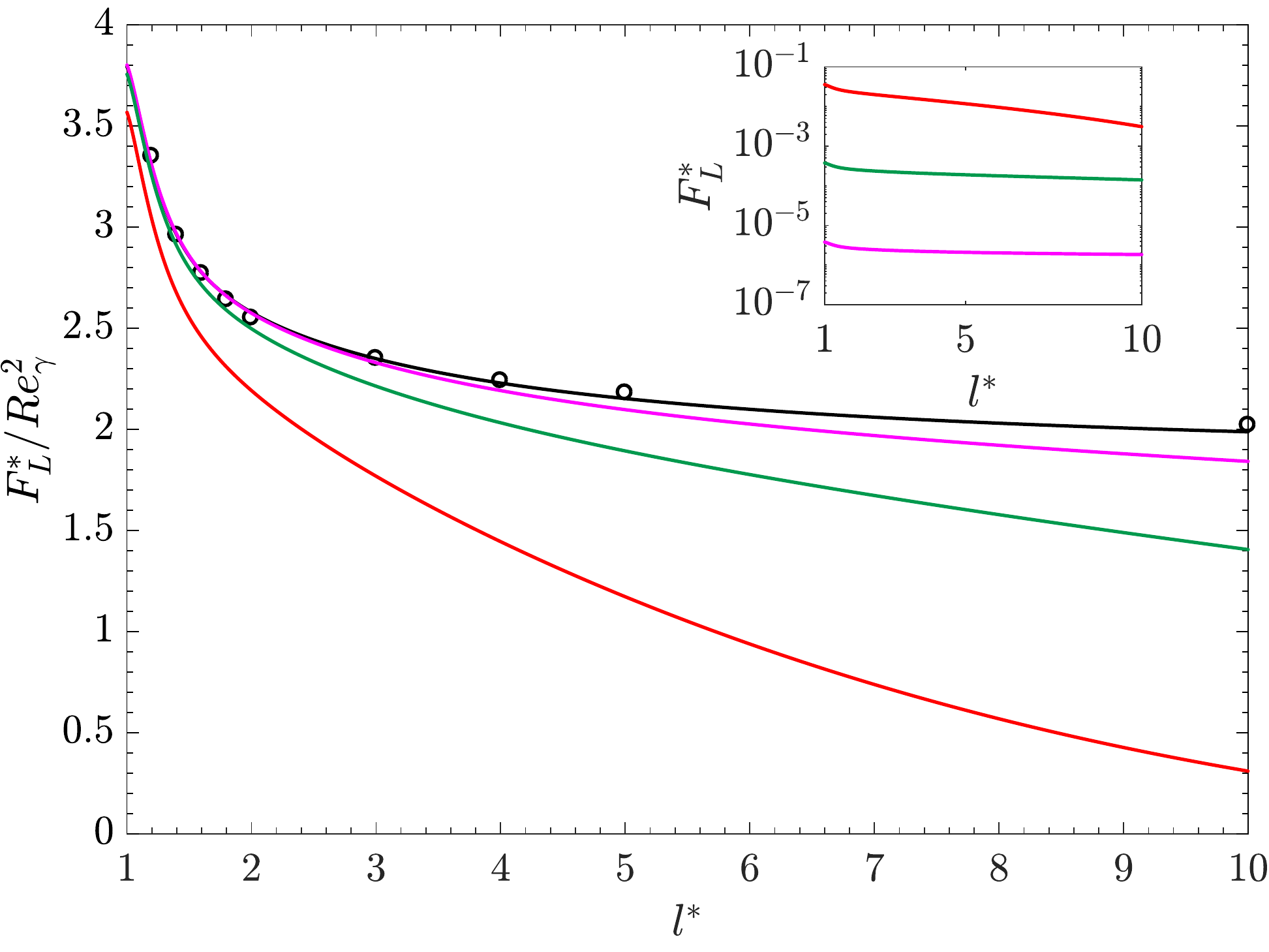}}
\caption{}
\label{netlift1}
\end{subfigure}
\caption{Analysis of a freely suspended neutrally-buoyant particle translating in a linear shear flow near a wall. (a) Non-dimensional slip velocity when $C_\text{D,1}$ is evaluated using the present numerical correlation, Eq. (\ref{eq:dragCD3}) for $Re_\gamma =0$ (\protect\blackline), $Re_\gamma=10^{-3}$ (\protect\magentaline), $Re_\gamma=10^{-2}$ (\protect\greenline), $Re_\gamma=10^{-1}$ (\protect\redline).  Dashed lines indicate results evaluated outside the strict range in which the new correlation has been validated.  Slip velocity when $C_\text{D,1}$ is evaluated using the \citet{Magnaudet1} correlation, Eq. (\ref{eq:dragCD1}) for $Re_\gamma = 0$ (\protect\blueline); \citet{Fischer87} numerical results for $Re_\gamma \ll 1$ ($\circ$); \citet{Goldman2} numerical results for Stokes flow ($\bullet$). (b-d) Different contributions to the net lift force acting on the particle,  ${F_\text{L,1}^{\prime}}, {F_\text{L,2}^{\prime}}$ and ${F_\text{L,3}^{\prime}}$ representing shear, slip-shear and slip effects, respectively. (e) Net lift force non-dimensionalised by shear Reynolds number.}
\label{slip}
\end{figure}

The slip velocity of a force free particle is first calculated by setting the net drag force (${F_\text{D}}^{\ast}$) in Eq. (\ref{eq:drag1}) to zero. The resulting non-dimensional velocity $u^{\ast}_\text{slip} (= Re_\text{slip}/Re_\gamma$) is:
\begin{equation}
u^{\ast}_\text{slip}=-\dfrac{C_\text{D,1}}{C_\text{D,2}}
\label{eq:dragCDfree}
\end{equation}
The present numerical drag correlation for $C_\text{D,1}$ provided by Eq. (\ref{eq:dragCD3}) is used to evaluate the slip velocity for three finite shear Reynolds numbers ($Re_\gamma = 10^{-1}$, $10^{-2}$, $10^{-3}$) as well as $Re_\gamma = 0$. In these calculations the inner region Faxen drag correlation given by \citet{Magnaudet1} (with higher order terms up to $\mathcal{O}(({1/l^{\ast}})^5)$) is used to evaluate $C_\text{D,2}$. This $C_\text{D,2}$ correlation has been experimentally validated by \cite{Ambari,Takemura} up to $Re_\text{slip} = 0.1$. The calculated $u^{\ast}_\text{slip}$ values are plotted as a function of separation distance in figure \ref{sliponly}. The velocities are compared with: i) numerical predictions by \citet{Fischer87} based on a Boundary Element Method (BEM) that included small inertial effects ($Re_\gamma \ll 1$); ii) theoretical/numerical predictions by \citet{Goldman2} applicable to Stokes flow; and iii) slip velocities again calculated using Eq. (\ref{eq:dragCDfree}), but with $C_\text{D,1}$ evaluated using the \citet{Magnaudet1} correlation of Eq. (\ref{eq:dragCD1}).

Consistent with earlier studies \citep{Goldman2, TakemuraMagnaudet2}, the negative slip velocities obtained confirm that a neutrally-buoyant particle translating in a linear shear flow lags the fluid when in close proximity to a wall. The present correlation, Eq. (\ref{eq:dragCD3}), used for evaluating $C_\text{D,1}$ predicts a slip velocity that is consistent with the two available numerical results down to reasonably small separation distances ($l^{\ast} \sim 1.1$).  In comparison the results obtained using the \citet{Magnaudet1} correlation (Eq. (\ref{eq:dragCD1})) diverge from the numerical results at $l^{\ast} \sim 1.5$.  At small separation distances ($l^{\ast} \lesssim 1.1$) the two numerical results diverge.  The predicted Stokes flow velocities from \citet{Goldman2} exhibit large negative values when the particle is almost in contact with the wall ($l^{\ast} \sim 1$), whereas the few results of \citet{Fischer87} in this region, that are valid for small but finite inertial effects, show smaller slip velocities. The difference between \citet{Fischer87} and \citet{Goldman2} may be due to insufficient numerical accuracy of a Gauss-Legendre product formula used by \citet{Fischer87}, as suggested by \citet{Shi2}.  We make no further comment on this difference, but note that the extrapolation of our correlation (indicted by the dashed lines in figure \ref{sliponly}) appears to be more consistent with the \citet{Fischer87} results at $l^{\ast} = 1.05$ than those of \citet{Goldman2}.

Slip velocity curves for the selected range of Reynolds numbers suggest that the shear rate has only a small effect ($\sim \mathcal{O}(10^{-3})$) on $u^{\ast}_\text{slip}$ for all separation distances. As expected, predicted slip velocities reduce to small values with increasing separation distance for the low $Re_\gamma = 10^{-3}, 10^{-2}$ cases. However, for $Re_\gamma = 10^{-1}$ the effect of inertia is relatively more important and the slip velocity further reduces, becoming approximately zero.  In fact, small positive slip velocities are predicted under the conditions $Re_\gamma = 10^{-1}$ and $l^{\ast} > 9.2$. However, as mentioned in Section \ref{sec:Drag force}, numerical inaccuracy of the fitted correlation, Eq. (\ref{eq:dragCD3}), at large separation distances for high $Re_\gamma$, is more likely to be the main reason that such positive slip velocities are obtained, rather than being physically meaningful.  Significantly, the slip velocities are very small in this high separation distance region for all $Re_\gamma$ considered.

\subsubsection{Lift force and migration velocity}

By using the slip velocity calculated via Eq. (\ref{eq:dragCDfree}), the lift force on a neutrally-buoyant particle experiencing both slip and shear within a linear shear flow can be written:
\begin{gather}
     \frac{F_{L}^{\ast}}{Re_{\gamma}^2} = C_\text{L,1}+C_\text{L,2} u^{\ast}_\text{slip}+C_\text{L,3} {u^{\ast}_\text{slip}}^2\\
     \intertext{which may also be written in the alternative form:}
     {F_\text{L,tot}^{\prime}} = {F_\text{L,1}^{\prime}}+{F_\text{L,2}^{\prime}}+{F_\text{L,3}^{\prime}}
\label{eq:liftfree}
\end{gather}
Here ${F_\text{L,tot}^{\prime}} (={F_\text{L}^{\ast}}/{Re_{\gamma}^2}$) is the net lift force non-dimensionalised by shear Reynolds number. The first term ${F_\text{L,1}^{\prime}}$ ($=C_\text{L,1}$), provides the wall lift force contribution due to shear only (as discussed in this paper), while the second and third terms, ${F_\text{L,2}^{\prime}} (=C_\text{L,2} u^{\ast}_\text{slip})$ and ${F_\text{L,3}^{\prime}}(=C_\text{L,3} {u^{\ast}_\text{slip}}^2)$, respectively, provide the lift force contributions due to shear combined with slip. These contributions, as well as the net lift force, are plotted in figure \ref{slip} for $Re_\gamma \ll 1$ and $Re_\gamma = 10^{-3}, 10^{-2}, 10^{-1}$. In all cases the lift force coefficients $C_\text{L,2}$ and $C_\text{L,3}$ are evaluated using the \citet{CherukatMcLaughlinCorrection} correlations, whereas $C_\text{L,1}$ is evaluated using our freely rotating particle correlation, as given by Eq. (\ref{0slipnewmodel}).

For separation distances $l^{\ast} \gtrsim 2$, the force contributions illustrate that ${F_\text{L,1}^{\prime}} > {F_\text{L,2}^{\prime}} \gg {F_\text{L,3}^{\prime}}$ for a freely translating particle. This indicates that the wall-shear lift contribution, ${C_\text{L,1}}$, either dominates or is similar to the other forces in this high force region, with the consequence that an accurate inertial correction for ${C_\text{L,1}}$ is required to accurately predict the net lift force acting on a freely translating particle. Figure \ref{netlift1} also shows that the estimated ${F_\text{L,tot}^{\prime}}$ using our correlations agrees well with the low $Re_\gamma$ results of \citet{Fischer87} (using $Re_\gamma = 0$ in our correlation), but deviates significantly from these results as the shear Reynolds number and separation distance increase.  This again reinforces the importance of the Reynolds number correction to ${C_\text{L,1}}$. Interestingly, the inset of figure \ref{netlift1} illustrates that the net lift force, ${F_\text{L}^{\ast}}$, while increasing with shear rate, is relatively independent of separation distance, except at the highest $Re_\gamma$ considered.

\begin{figure}
\begin{subfigure}{\linewidth}
\centerline{\includegraphics[width=0.8\textwidth]{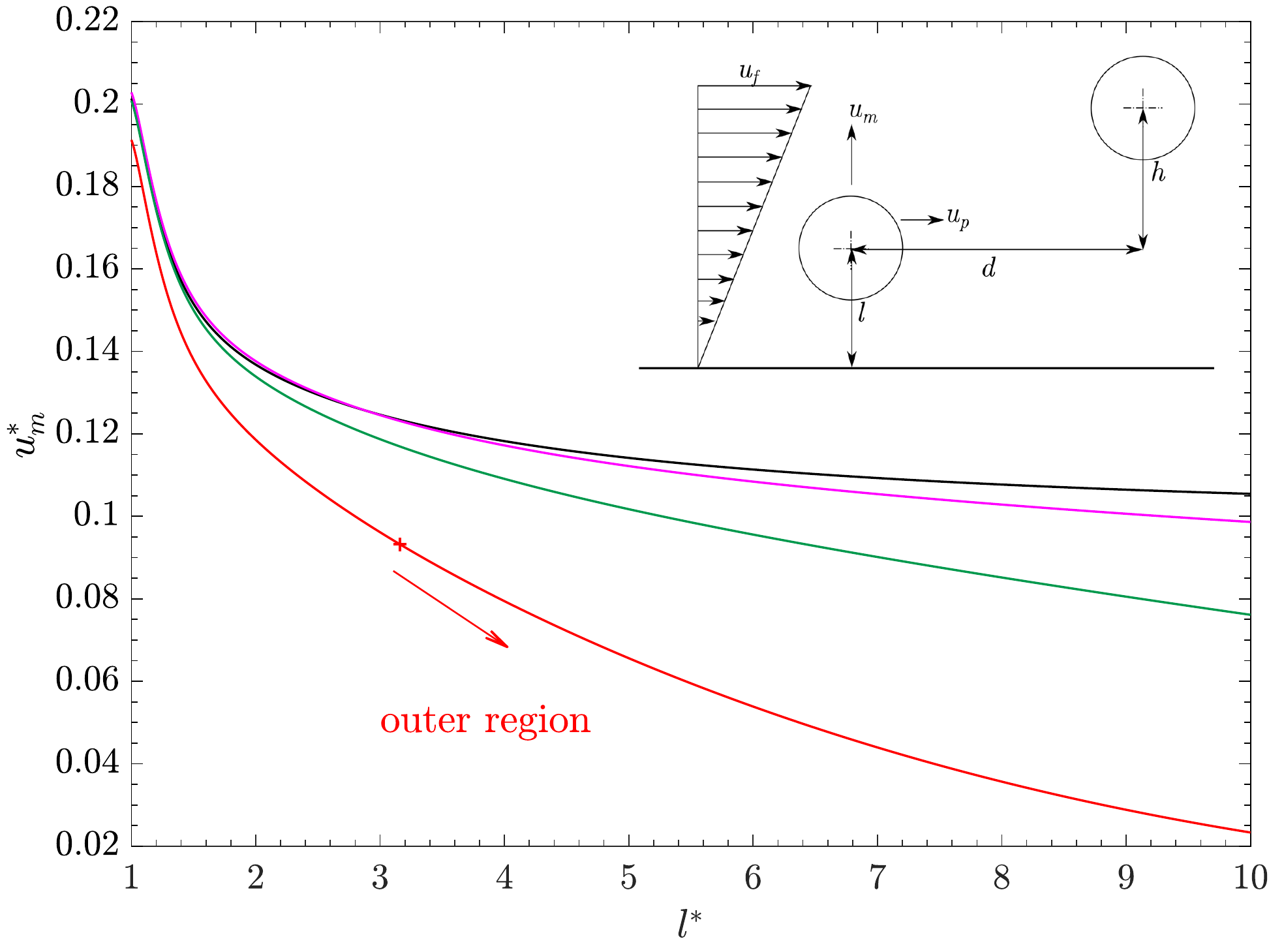}}
\caption{Dimensionless migration velocity as a function of dimensionless particle location $l^{\ast}$}
\label{MigrationVelocityNondim}
\end{subfigure}
\begin{subfigure}{\linewidth}
\centerline{\includegraphics[width=0.8\textwidth]{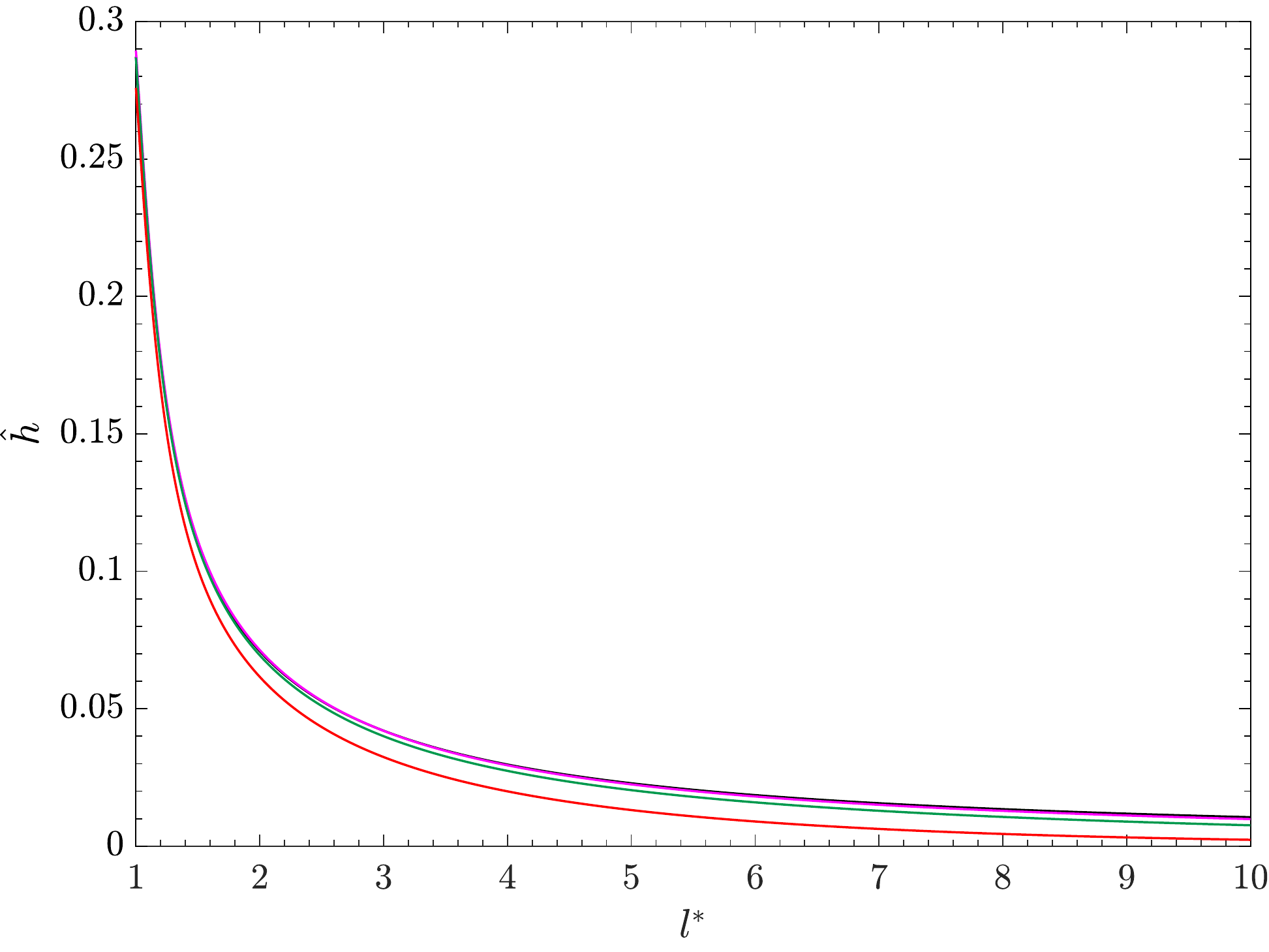}}
\caption{Lateral displacement per unit movement in flow direction for a given time as a function of dimensionless particle location $l^{\ast}$.}
\label{DisplacementNondim}
\end{subfigure}
\caption{Migration velocity and displacement of a freely suspending neutrally-buoyant particle for $Re_\gamma \ll 1$ (\protect\blackline), $Re_\gamma=10^{-3}$ (\protect\magentaline), $Re_\gamma=10^{-2}$ (\protect\Limegreenline) and $Re_\gamma=10^{-1}$ (\protect\redline).}
\label{disvelo}
\end{figure}

Dimensionless migration velocities $u_\text{mig}^{\ast}$, obtained by dividing ${F_\text{L,tot}^{\prime}}$ by $6\pi$, are presented in Figure \ref{MigrationVelocityNondim} for the same set of shear Reynolds numbers. Based on the migration velocity, the lateral displacement per unit longitudinal displacement  ${\hat{h}} (=h/d)$ of a particle located at $l^{\ast}$ is also given in figure \ref{DisplacementNondim}. For all shear Reynolds numbers the lateral displacement of a freely translating particle closer to the wall is much higher than that of a particle travelling further away from the wall. Indeed, the maximum lateral displacement $\hat{h}\sim 0.28$ that occurs when the particle is almost touching the wall would quite quickly push a free particle away from the wall. The lateral displacement also displays a weak dependence on the shear rate close to the wall ($l^{\ast} \sim 1$), however when a particle is away from the wall $\hat{h}$ decreases significantly as the shear Reynolds number increases.

\section{Conclusion}\label{Conclusion}

The lift and drag forces acting on a spherical particle in a single wall bounded linear shear flow field are examined via numerical computation. Forces are obtained under a zero slip condition, by setting the particle to translate at the same velocity as fluid. The Navier-Stokes equations are solved using a finite-volume solver to find the fluid flow around the particle.  The effect of shear rate and wall presence are investigated by varying the shear Reynolds number over the range $Re_\gamma = 10^{-3} - 10^{-1}$, and the wall separation distance over ${l^{\ast}} = 1.2 - 9.5$. Sensistivity analyses show that large computational flow domains, combined with careful mesh construction, are required to accurately predict these forces at these low shear rates and wall separations. Computed lift and drag coefficients at $Re_\gamma \sim \mathcal{O}(1)$ are compared against theoretical values, predicted by asymptotic models mainly derived for $Re_\gamma \ll 1$. Accounting for the variations, slip independent inertial corrections are suggested for both the lift and drag force coefficients.

The computed lift force coefficients at the lowest Reynolds numbers are in good agreement with the previous theoretical results when the particle is close to the wall \citep{CherukatMcLaughlin,CoxHsu,Krishnan}. With increasing shear rate, a significant decrease of the computed lift coefficient is observed for both non-rotating and freely rotating particles. This decrease of lift coefficient is further enhanced when the wall is away from the particle and is located in the outer region. Two numerical lift correlations for non-rotating and freely rotating conditions are presented, accounting for both the inner and outer region behaviour in the limit of finite $(Re_\gamma)$. These expressions reduce to theoretical values predicted in the three limits of $Re \rightarrow 0, {l^{\ast}} \rightarrow 1$ and ${l^{\ast}} \rightarrow \infty$.

Drag force coefficients computed for the freely-rotating sphere agree reasonably well with low $(Re_\gamma)$ results from a previous analytical study over most of the wall separation range \citep{Magnaudet1}, however considerable deviation is seen when the particle is within one radius of the wall.  Consequently a drag coefficient model is proposed that includes higher order terms in separation distance (up to $\mathcal{O}((1/{l^{\ast}})^6)$) that more accurately captures this near-wall drag behaviour. A shear based inertial correction, independent of slip velocity, is also suggested for the modified drag coefficient.

The behaviour of a neutrally-buoyant particle suspended in a linear shear flow is analysed using the new drag and lift correlations. Slip velocities calculated within the range of validity of our correlations compare favourably with previously presented results that are valid for low $(Re_\gamma)$. The negative slip velocities indicate that neutrally-buoyant particles lag the fluid near walls for low $(Re_\gamma)$. The lift force of a neutrally-buoyant particle at finite $(Re_\gamma)$, accounting for both shear induced slip and slip induced lift, is also analysed. A rapid decrease of the lift force is observed as both the shear Reynolds number and separation distance increase. This behaviour is not captured via existing analytical lift correlations that are limited to $Re \ll 1$.
Overall, the suggested correlations will aid in providing accurate constitutive equations for interphase forces to predict the behaviour of neutrally-buoyant particles moving near walls.

\section{Acknowledgements}
Support from the Australian Research Council (LP160100786) and CSL is gratefully acknowledged. One of the authors (NE) acknowledges the support from the Melbourne Research Scholarships program of The Melbourne University. Declaration of Interests: The authors report no conflict of interest.

\appendix
\section{}\label{appA}
\renewcommand{\thefigure}{\Alph{section}.\arabic{figure}}
\setcounter{figure}{0}
\FloatBarrier
\begin{minipage}{\linewidth}
    \centering
    \includegraphics[width=1.0\textwidth]{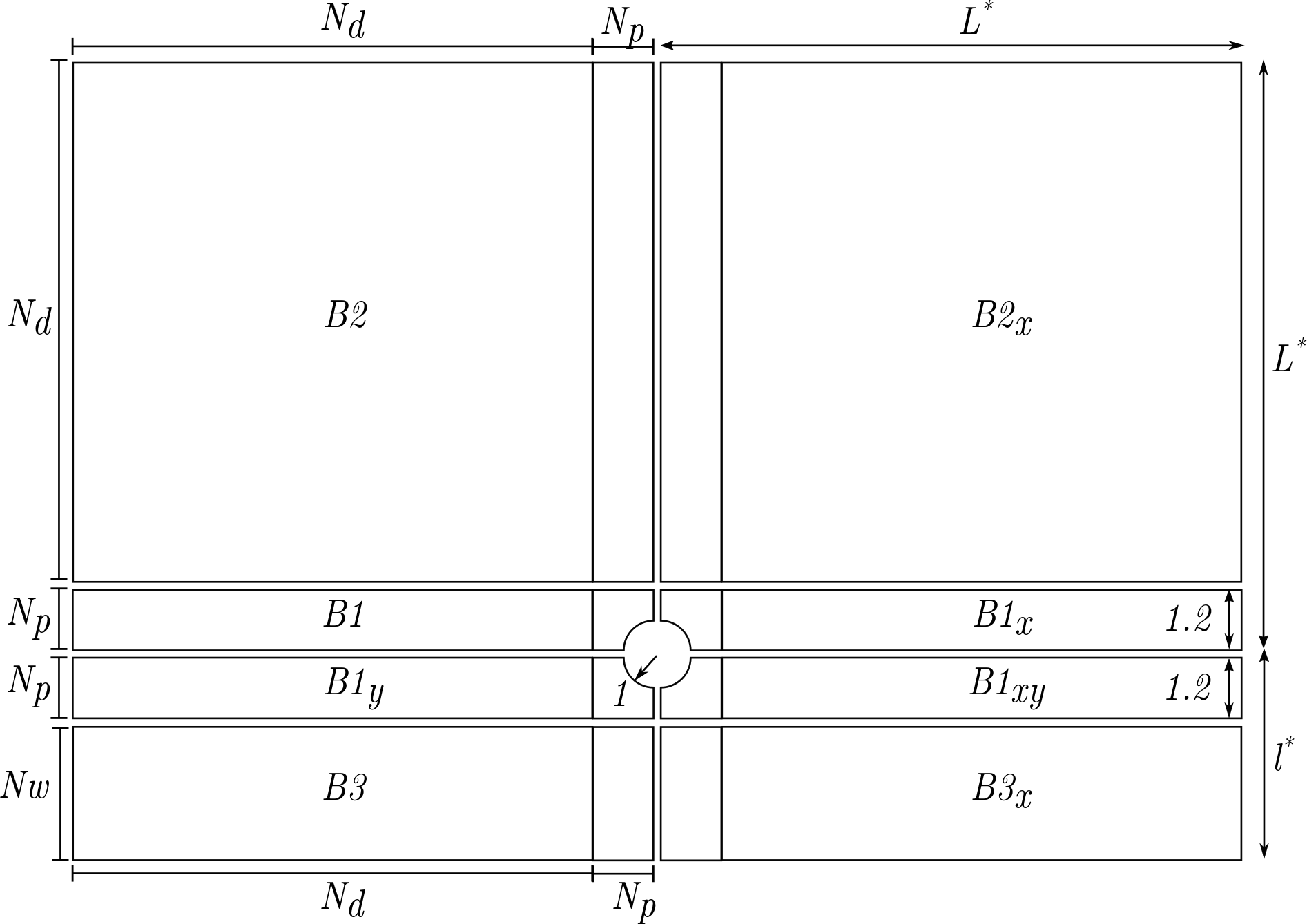}
    \captionof{figure}{A detailed 2D frame at $z=0$ of ``lego" mesh showing how individual blocks are duplicated to form the complete mesh. $N_\text{d}, N_\text{p}$ and $N_\text{w}$ indicate number of points, while $L^{\ast}, l^{\ast}$ are dimensionless domain sizes.}
    \label{BlockMesh}
\end{minipage}
\bibliographystyle{jfm}
\bibliography{ekanayake19b_arxiv}
\end{document}